\documentclass[12pt]{article}
\usepackage{axodraw,a4,epsfig}

\newcommand{\be}{\begin{equation}}
\newcommand{\ee}{\end{equation}}

\newcommand{\beq}{\begin{equation}}
\newcommand{\eeq}{\end{equation}}
\newcommand{\beqa}{\begin{eqnarray}}
\newcommand{\eeqa}{\end{eqnarray}}
\def\al{\alpha}
\newcommand{\Eqn}[1]{Eq.~(\ref{#1})}
\newcommand{\Eqns}[2]{Eqs.~(\ref{#1}) and (\ref{#2})}

\newcommand{\cL}{{\cal L}}

\newcommand{\e}{\epsilon}

\newcommand{\RS}{\rm RS}
\newcommand{\PS}{\rm PS}

\newcommand{\MS}{\overline{\rm MS}}
\def\als{\alpha_{\rm s}}
\newcommand{\nn}{\nonumber}

\begin{document}

\pagestyle{empty}
\begin{flushright}
  IPPP/06/51 \\
  UB-ECM-PF-06-18
\end{flushright}
\vspace*{1cm}
\begin{center}
  {\sc \large Heavy Quark Pair Production near Threshold with\\[5pt]
  Potential Non-Relativistic QCD} \\
   \vspace*{2cm} {\bf Antonio~Pineda\footnote{Permanent address after
September 1st: Grup de F\'\i sica Te\`orica and IFAE, Universitat
Aut\`onoma de Barcelona, E-08193 Bellaterra, Barcelona, Spain.}$^a$
and Adrian~Signer$^b$}\\
\vspace{0.6cm}
{\it
$^a$\ Dept. d'Estructura i Constituents de la Mat\`eria\\
U. Barcelona, Diagonal 647, E-08028 Barcelona, Catalonia, Spain\\[10pt]
$^b$\ Institute for Particle Physics Phenomenology \\
Durham, DH1 3LE, England \\}
  \vspace*{2.4cm}
  {\bf Abstract} \\
  \end{center}
We study the effect of the resummation of logarithms for $t\bar{t}$
production near threshold and inclusive electromagnetic decays of
heavy quarkonium.  This analysis is complete at
next-to-next-to-leading order and includes the full resummation of
logarithms at next-to-leading-logarithmic accuracy and some partial
contributions at next-to-next-to-leading logarithmic accuracy.
Compared with fixed-order computations at next-to-next-to-leading
order the scale dependence and convergence of the perturbative series
is greatly improved for both the position of the peak and the
normalization of the total cross section. Nevertheless, we identify a
possible source of large scale dependence in the result. At present we
estimate the remaining theoretical uncertainty of the normalization of
the total cross section to be of the order of 10\% and for the
position of the peak of the order of 100~MeV.

\vspace*{5mm}
\noindent

\newpage

\setcounter{page}{1}
\pagestyle{plain}

\section{Introduction}

The existence of (heavy) quarks with a large mass compared with
$\Lambda_{QCD}$, like the top, the bottom, and maybe the charm, makes
particularly interesting the study of physical processes where a heavy
quark pair is created close to threshold, because accurate
experimental information over the whole near-threshold region may
allow for a precise determination of some parameters of the Standard
Model. For instance, the future International Linear Collider (ILC)
offers the opportunity to study the top quark with unprecedented
accuracy \cite{FadinKhoze,Martinez:2002st,Hoang:2000yr}. To fully
exploit the potential of an ILC in this respect, it is essential that
a dedicated measurement of the cross section for the production of a
top antitop quark pair close to threshold is made. Such a threshold
scan allows for an extremely precise measurement of the top-quark mass
and yields information on the top-quark width and the top-Higgs Yukawa
coupling.  The analogous threshold scan for the $b \bar{b}$ sector is
also fundamental for non-relativistic sum rules
\cite{bottomSR,Melnikov2,Penin,Hoang:1999ye,Beneke:1999fe} and may lead to
accurate determinations of the bottom quark mass \cite{Pineda:2006gx}.

The characteristic feature of heavy-quark pair production close to
threshold is the smallness of the relative velocity $v$ of the heavy
quarks in the centre of mass frame. This entails a hierarchy of scales
$\mu_h\gg \mu_s\gg \mu_{us}$ where the hard scale $\mu_h$ is of the
order of the heavy quark mass $m$, the soft scale $\mu_s\sim m v$ is
of the order of the typical momentum of the heavy quarks and the
ultrasoft scale $\mu_{us}\sim m v^2$ is of the order of the typical
kinetic energy of the heavy quarks. The presence of an additional
small parameter can be exploited by systematically expanding in the
strong coupling $\alpha_s$ and $v$. Thus, in this context, a
next-to-leading order (NLO) calculation takes into account all terms
that are suppressed by either $\alpha_s$ or $v$ relative to the
leading-order (LO) result, whereas a next-to-next-to-leading order
(NNLO) calculation includes all terms suppressed by two powers of the
small parameter $\alpha_s\sim v$. The coefficients of this
perturbative series contain large logarithms $\log v$. In order to
improve the reliability of the calculation, these logarithms should be
resummed. Counting $\alpha\log v \sim 1$ in a NLO or NNLO calculation
produces a result of next-to-leading logarithmic (NLL) or
next-to-next-to-leading logarithmic (NNLL) accuracy
respectively. There are no leading logarithmic (LL) corrections, thus
the LO and LL results are the same.

The expansion as well as the resummation of the logarithms can be
organized most efficiently by using an effective theory (for a review
see Ref.~\cite{Brambilla:2004jw}) approach. This is done most
conveniently by using the threshold expansion~\cite{Beneke:1997zp}
that allows to separate the full result of an integral into
contributions due to the various modes. Denoting the generic
integration momentum by $k = (k^0, {\bf k})$, the modes that are
relevant are: the hard mode $k^0\sim{\bf k}\sim \mu_h$, the soft mode
$k^0\sim{\bf k}\sim \mu_s$, the potential mode $k^0\sim \mu_{us}, {\bf
k}\sim \mu_s$ and the ultrasoft mode $k^0\sim{\bf k}\sim
\mu_{us}$. The standard procedure is to first match QCD to
non-relativistic QCD (NRQCD)~\cite{nrqcd} at the hard scale. This
corresponds to integrating out the hard modes using the threshold
expansion. The resulting theory is then matched to potential NRQCD
(pNRQCD)~\cite{pnrqcd,Brambilla:1999xf} by integrating out the soft
modes and potential gluon modes. At this stage the theory consists of
a non-relativistic quark pair interacting through potentials and
ultrasoft gluons.

Within this framework NNLO calculations have been performed by several
groups (for a review in the case of the top quark see
\cite{Hoang:2000yr}) and quite a few partial results needed for a
NNNLO calculation have been obtained \cite{Brambilla:1999qa,
Kniehl:1999ud, Brambilla:1999xj, Kniehl:2001ju, Penin:2005eu,
Beneke:2005hg, Marquard:2006qi}.  Moreover, the existence of the
effective theory also provides the necessary framework on which to use
renormalization group (RG) techniques.  These allow to resum the large
logarithms, $\log v$, that appear as the ratio of the different scales
appearing in the physical system: $\log \mu_h/\mu_s$ and $\log
\mu_s/\mu_{us}$.  At present, the situation is as follows. The
computation of the heavy quarkonium spectrum is known with NNLL
accuracy \cite{Pineda:2001ra,Hoang:2002yy} (for the hyperfine
splitting at NNNLL \cite{Kniehl:2003ap, Penin:2004xi}).  Inclusive
electromagnetic decays of heavy quarkonium have been computed to NLL
\cite{Pineda:2001et,Hoang:2002yy} and for the spin-zero, spin-one
ratio to NNLL \cite{Penin:2004ay}. In the case of top-quark pair
production a renormalization-group improved (RGI) calculation is
available~\cite{Hoang:2000ib, Hoang:2001mm}, using a somewhat
different approach, referred to as vNRQCD~\cite{vnrqcd} (in this
theory, soft degrees of freedom are kept dynamical and the matching
from QCD to vNRQCD is carried out directly).  However, so far, there
is no RGI calculation of heavy-quark pair production within the
conventional pNRQCD approach. It is the main purpose of this work to
close this gap and to provide the ingredients to perform such
computations in the case of top and bottom-quark pair production. We
will report results on $t\bar t$ production near threshold, as well as
elaborate on the computation of non-relativistic sum rules presented
in Ref.~\cite{Pineda:2006gx}.  Inclusive electromagnetic decays of
heavy quarkonium will also be considered.  These analyses will be
complete at NNLO and include the complete resummation of logarithms at
NLL accuracy and some partial contributions at NNLL.

The importance of higher order logarithms can be illustrated in the
top quark case. In the on-shell scheme, the NNLO corrections turned
out to be much larger than anticipated and, moreover, made the
theoretical prediction very strongly scale dependent.  If the cross
section is expressed in terms of a threshold mass~\cite{massdef,
Beneke:1998rk, Pineda:2001zq} rather than the pole mass, the position
of its peak is more stable and can be predicted with a small theoretical
error. This will allow to determine the top threshold mass and
ultimately the top $\MS$-mass with a very small error. The situation
is much less favourable regarding the normalization of the cross
section. The corrections are huge and the scale dependence at NNLO is
larger than at NLO, indicating that this quantity is not well under
control at NNLO. This affects how accurately one may obtain the
top-quark width and the top-Higgs Yukawa coupling.  It has been shown
that the inclusion of the potentially large $\log v$ terms is
numerically very important and improves the situation regarding the
normalization of the cross section considerably~\cite{Hoang:2000ib,
Hoang:2001mm}. One of the aims of this paper is to compare with these
results. We also investigate to what extent the resummation of the
logarithms is really required and to what extent the improvement we
observe in the RGI results is simply due to the partial inclusion of
higher-order terms. To do so we will produce ``NNNLO'' results by
re-expanding full RGI results and dropping terms that are beyond
NNNLO. Of course, these results are by no means complete at NNNLO, but
they contain those NNNLO terms that are enhanced by a logarithm. To
obtain an estimate of the importance of resummation, we then compare
these ``NNNLO'' curves to the full RGI results. As we will see, this
partial inclusion of NNNLO terms does reproduce the bulk of the RGI
result except for rather small values of $\mu_s$. In particular, in
the case of the top quark the difference between fully resummed and
partial NNNLO results is small.

The paper is structured as follows: In Section 2 we present our
formalism and describe how to perform a RGI calculations in
pNRQCD. Some formulae and technical details of this section are
relegated to the Appendix. In Section 3 we apply these results to
$t\bar t$ production near threshold. Section 4 contains the
application to bottomonium non-relativistic sum rules and inclusive
electromagnetic decay widths. In the final section, we show our
conclusions.

\section{Effective theory}

Within pNRQCD, $\psi$ and $\chi$, the fields representing the
non-relativistic quark and antiquark, interact through a potential and
with dynamical ultrasoft gluons. It is well known that the leading
Coulomb interaction is not suppressed and has to be included in the
LO Lagrangian which is given by
\begin{eqnarray}
\cL^{(0)}_{\rm pNRQCD} &=& 
\psi^\dagger \left(i\partial^0 + \frac{\partial^2}{2m} \right)\psi + 
\chi^\dagger \left(i\partial^0 - \frac{\partial^2}{2m} \right)\chi
\label{L0pnrqcd}
\\
&+& \int d^3 {\bf r} \left(\psi^\dagger T^a \psi\right)
\left(-\frac{\alpha_s}{r}\right) 
\left(\chi^\dagger T^a \chi\right)
\,,
\nonumber
\end{eqnarray}
where $T^a$ are the colour matrices and the strong coupling is
understood to be evaluated at the soft scale, $\alpha_s\equiv
\alpha_s(\mu_s)$, unless explicitly indicated otherwise. Subleading
effects are incorporated in the Lagrangian as corrections to the
potential, $\delta V$, and as interactions of the heavy quarks with
ultrasoft gluons. For further details we refer to
Ref.~\cite{Brambilla:2004jw}. If we restrict the accuracy of our
analysis to NNLL, ultrasoft gluons do not appear as physical final
states (thought their effect is embedded in the RGI running of the
matching coefficients of the potentials and currents). It follows that
the number of particles is conserved (we only have one heavy quark and
one heavy antiquark) and the problem effectively becomes equivalent to
do standard quantum mechanics perturbation theory.  If we restrict
ourselves to study the Hilbert space spanned by the heavy
quark-antiquark system in the singlet colour sector we are lead to
solve the following equation for the associated Green function
\begin{equation}
(H({\bf r},{\bf p})-E)G({\bf r},{\bf r}^{\, \prime};E)=
\delta({\bf r}-{\bf r}^{\, \prime})
\,,
\end{equation}
where 
\begin{equation}
H=H_c+\delta V
\end{equation} 
and 
\begin{equation}
H_c=\frac{{\bf p}^2}{m}-C_F\frac{\als}{r}
\,,
\end{equation}
with $C_F=(N_c^2-1)/(2N_c)$ (where the colour factor $N_c=3$).  The
explicit expression of $\delta V$ (the correction to the Coulomb
potential) will be given afterwards.  The full Green function $G$ can
be solved iteratively in an expansion in the velocity by performing
multiple insertions of $\delta V$.  The LO solution is the Coulomb
Green function $G_c$ and we write
\begin{equation}
G({\bf r},{\bf r}^{\, \prime};E)=G_c({\bf r},{\bf r}^{\, \prime};E)
+\delta G({\bf r},{\bf r}^{\, \prime};E)\,,
\end{equation}
where 
\begin{equation}
\label{GcSym}
G_c({\bf r},{\bf r}^{\, \prime};E)=
\langle {\bf r}|\frac{1}{H_c-E}|{\bf r}^{\, \prime}\rangle
\end{equation}
and 
\begin{equation}
\label{pertth}
\delta G({\bf r},{\bf r}^{\, \prime};E)=
-\langle {\bf r}|\frac{1}{H_c-E}\delta V 
\frac{1}{H_c-E}|{\bf r}^{\, \prime}\rangle
+
\cdots
.
\end{equation}

$G({\bf r},{\bf r}^{\, \prime};E)$ is related to the correlators that
appear in the total cross section for the production of a heavy quark
pair, $\sigma(e^+e^- \to Q\bar{Q})$ with the centre of mass energy
$\sqrt{q^2}=\sqrt{s}\sim 2 m$. This is the key quantity we are
interested in. The cross section obtains contributions from $\gamma$
and $Z$ exchange. In order to simplify the discussion we ignore the
$Z$ exchange in what follows. The cross section can then be written as
\begin{equation}
\sigma^\gamma (s) = \frac{4 \pi \alpha^2_{\rm EM}}{3\, s}\, e_Q^2\, R(s)
\,,
\label{sigma}
\end{equation}
where $e_Q$ is the electric charge of the heavy quark, $\alpha_{\rm
EM}$ the electromagnetic coupling at the hard scale and the ratio
$R\equiv \sigma(e^+e^-\to Q\bar Q) / \sigma(e^+e^-\to \mu^+\mu^-)$ is
expressed as a correlator of two heavy-quark vector currents $j^\mu(x)
\equiv \bar{Q} \gamma^\mu Q(x)$
\begin{equation}
R(s) = \frac{4\pi}{s}\,  {\rm Im} \left(-i \int d^4 x\,  e^{i q\cdot x} 
\langle 0 | T\{j^\mu(x)\, j_\mu(0)\} | 0\rangle \right)
\,.
\label{Rcorr}
\end{equation}
In order to compute this correlator, we first express the current in
terms of the non-relativistic two-component spinor fields $\psi^{\dagger}$ 
and $\chi$,
\begin{equation}
{\bar Q} \gamma^{\mu} Q = 
c_1\chi^{\dagger}\sigma^i\psi -
\frac{d_1}{6m^2}\chi^{\dagger}\sigma^i\left(i {\bf D} \right)^2 \psi +
\ldots\ ,
\label{NRQCDcurrent}
\end{equation}
where the matching coefficients $c_1$ and $d_1$ are normalized to 1 at
LO. By using the equations of motion, \Eqn{NRQCDcurrent}
can also be written in the following way
\begin{equation}
{\bar Q} \gamma^{\mu} Q = c_1\chi^{\dagger}\sigma^i\psi
- {d_1 \over 6m}i \partial_0 
\left( \chi^{\dagger}\sigma^i\psi \right) + \cdots
\,.
\label{nrcurrent}
\end{equation}

Given the Lagrangian, we solve the corresponding Schr\"odinger
equation and ultimately relate the imaginary part of the spin one
Green function at the origin to $R(s)$ by the equality
\begin{equation}
\label{RE}
R(E)=\frac{24\pi e_Q^2N_c}{s}
\left(c_1^2-c_1d_1\frac{E}{3m}\right)
{\rm Im}\,G_{s=1}({\bf 0},{\bf 0};E)
\,,
\end{equation}
which is valid with NNLL accuracy (where $E\equiv\sqrt{s}-2m$). We
note that in evaluating $R(E)$ we expand the expression in \Eqn{RE}
and drop all terms that are beyond NNLO/NNLL. In particular, we set
$c_1=1$ in the second term of the parenthesis.

In order to connect with the notation in
Ref.~\cite{Brambilla:2004jw}, \Eqn{RE} can also be written as
\begin{equation}
\label{RErmp}
R(E)=\frac{18 N_c}{m^2\alpha^2_{\rm EM}}
{\rm Im}\,G_{s=1}({\bf 0}, {\bf 0};E)
\left[
{\rm Im} f_{\rm EM}^{\rm pNR}({}^3S_1)
+{\rm Im} g_{\rm EM}^{\rm pNR}({}^3S_1){E \over m}
\right]
\,,
\end{equation}
with
\begin{eqnarray}
{\rm Im} f_{\rm EM}^{\rm pNR}({}^3S_1)&=&
{\pi e_Q^2\al_{\rm EM}^2 \over 3}c_1^2
\,,
\nn
\\
{\rm Im} g_{\rm EM}^{\rm pNR}({}^3S_1)&=&-{\pi e_Q^2\al_{\rm EM}^2 \over 3}
c_1\left(c_1+{1 \over3}d_1\right)
\,.
\end{eqnarray}

\subsection{Potential}

The computation of $G({\bf 0},{\bf 0};E)$ will lead to divergences.
These divergences are regularized by performing all calculations in
momentum space~\cite{Beneke:1999qg} and using dimensional
regularisation in $D=d+1=4-2\e$ dimensions. Thus, the LO
Green function at the origin, \Eqn{GcSym}, is understood as
\begin{equation}
G_c({\bf r},{\bf r}^{\, \prime};E) \Big|_{{\bf r}={\bf r}^{\, \prime}={\bf 0}}
\equiv \int \frac{d^d {\bf p}}{(2\pi)^d}
\frac{d^d {\bf p}^{\, \prime}}{(2\pi)^d}\, 
\tilde{G}_c({\bf p},{\bf p}^{\, \prime};E)
\,,
\label{greenorig}
\end{equation}
where $\tilde{G}$ denotes the Fourier transform of the Green function
and the insertions, \Eqn{pertth}, are to be evaluated as indicated in
\Eqns{singleIns}{doubleIns}.  Using the threshold
expansion~\cite{Beneke:1997zp}, the Fourier transform of the leading
order Green function can be computed as the sum of all ladder diagrams
with the exchange of potential gluons between the heavy quarks and can
be written as
\begin{eqnarray}
\tilde{G}_c({\bf p},{\bf p}^{\, \prime};E) &=& 
  (2\pi)^d \delta^{(d)}\left({\bf p}-{\bf p}^{\, \prime}\right) 
   \frac{-1}{E-{\bf p}^2/m}
\label{greenmom}
\\
&+& \frac{4 \pi C_F \alpha_s}{(E -{\bf p}^2/m) 
     \left({\bf p}-{\bf p}^{\, \prime}\right)^2 (E -{\bf p}^{\prime\, 2}/m)}
+ {\rm finite}
\,,
\nonumber
\end{eqnarray}
where we have omitted terms that are finite if \Eqn{greenmom} is used
in \Eqn{greenorig}. The Green function at the origin has a ultraviolet
divergence that manifests itself as a pole $1/\e$. Using
$\overline{\rm MS}$ subtraction and then taking the limit $\e\to 0$ we
find~\cite{Beneke:1999qg, Beneke:1999zr}
\begin{equation}
\label{Greenorig}
G_c({\bf 0},{\bf 0};E) = - \frac{\alpha_s\, C_F\, m^2}{4\pi} 
\left( \frac{1}{2\lambda} + \frac{1}{2} \log \frac{-4 m E}{\mu_s^2}
       - \frac{1}{2} + \gamma_E + \psi(1-\lambda) \right)
\end{equation}
where $\lambda\equiv C_F\, \alpha_s/(2 \sqrt{-E/m})$.  

Higher-order corrections to the Green function can be computed
perturbatively in two steps. First, the pNRQCD Lagrangian has to be
determined to the accuracy needed for the calculation. Second,
higher-order corrections are computed using quantum mechanics
perturbation theory, \Eqn{pertth}.  Since some of the potentials
generate singularities in insertions, we have to start from the NNLO
potential computed in $D$ dimensions~\cite{Beneke:1999qg}. It is
essential to manipulate the potential consistently in $D$ dimensions,
since it allows us to use the same hard matching coefficients defined
in the $\overline{\rm MS}$-scheme obtained in Refs.~\cite{c1nnlo,
Czarnecki:2001zc}. We
then bring the potential in a form that is more suitable to be
combined with the RGI coefficients as presented in
Ref.~\cite{Pineda:2001ra}. In particular we use
\begin{eqnarray}
\lefteqn{
\frac{C_F \pi \alpha_s}{m^2} 
\int \prod_{i=1}^4 \frac{d^d{\bf p}_i}{(2\pi)^d}\ 
\tilde G_c({\bf p}_1,{\bf p}_2) 
\left(\frac{{\bf p}_2^2-{\bf p}_3^2}{({\bf p}_2-{\bf p}_3)^2}\right)^2 
\tilde G_c({\bf p}_3,{\bf p}_4)} && 
\label{l2relation}
\\
& & \qquad  = \
\frac{C_F^2}{2}(1-2\e)
\frac{e^{\e \gamma_E} \Gamma^2(\frac{1}{2}-\e)\Gamma(\frac{1}{2}+\e)}
     {\pi^{3/2}\, \Gamma(1-2\e)} \ \times 
\nonumber \\
& & \qquad \qquad
\int \prod_{i=1}^4 \frac{d^d{\bf p}_i}{(2\pi)^d}\ 
\tilde G_c({\bf p}_1,{\bf p}_2) \,
\frac{\pi^2 \alpha_s^2\, \mu_s^{2\e}}{m\, |{\bf p}_2-{\bf p}_3|^{1+2\e}} 
\, \tilde G_c({\bf p}_3,{\bf p}_4)
\nonumber
\end{eqnarray}
to eliminate the $C_F^2$ term of the non-analytic potential
$1/q^{1+2\e}$, present in Ref.~\cite{Beneke:1999qg}.  The angular
momentum operator is generalized to
\begin{equation}
\frac{{\bf L}^2}{2\pi r^3} \to 
\left(\frac{{\bf p}^2-{\bf p}^{\prime\,2}}{{\bf q}^2}\right)^2 - 1
\label{lddef}
\end{equation}
to be compatible with $D$ dimensional calculations in momentum
space. Note that an insertion of this operator does not vanish even
for an $S$-wave (see \Eqn{AppAAMom}), but the corresponding
contribution could be absorbed into a redefinition of 
the matching coefficient of the current.

At NNLL, the higher-order corrections to the $D$-dimensional potential
in momentum space, $\delta\tilde V$, can then be written as
\begin{eqnarray}
\delta \tilde V &=&
  -\,  c_4 \frac{{\bf p}^4}{4m^3}\, (2\pi)^d\delta^{(d)}({\bf q})
- 4\pi C_F \frac{\alpha_{{\tilde V}_{s}}}{{\bf q}^2}+
4\pi C_F \frac{\als}{{\bf q}^2}
\label{Hpnrqcd}
\\
&&
-\,  C_F C_A D^{(1)}_s \frac{\pi^2\, \mu_s^{2\e}}{m\, q^{1+2\e}}\, (1-\e)
  \frac{e^{\e \gamma_E} \Gamma^2(\frac{1}{2}-\e)\Gamma(\frac{1}{2}+\e)}
        {\pi^{3/2}\Gamma(1-2\e)}
\nonumber \\
&&
-\,  \frac{2\pi C_F D^{(2)}_{1,s}}{m^2} 
 \frac{{\bf p}^2+{\bf p}^{\prime\, 2}}{{\bf q}^2}
+ \frac{\pi C_F D^{(2)}_{2,s}}{m^2} \left(
 \left(\frac{{\bf p}^2-{\bf p}^{\prime\, 2}}{{\bf q}^2}\right)^2 - 1 \right)
\nonumber \\
&& 
+\,  \frac{3\pi C_F D^{(2)}_{d,s}}{m^2}
- \frac{4\pi C_F D^{(2)}_{S^2,s}}{d\, m^2}\, 
       [{\bf S}_1^i,{\bf S}_1^j][{\bf S}_2^i,{\bf S}_2^j]
\nonumber \\
&& 
+\,  \frac{4\pi C_F D^{(2)}_{S_{12},s}}{d\, m^2}\, 
        [{\bf S}_1^i,{\bf S}_1^r][{\bf S}_2^i,{\bf S}_2^j]
   \left( \delta^{rj}-d\,  \frac{q^r q^j}{q^2} \right)
\nonumber \\
& & 
\nonumber
-\,   \frac{6\pi C_F D^{(2)}_{LS,s}}{m^2} \,
   \frac{p^i q^j}{q^2}
   \left([{\bf S}_1^i,{\bf S}_1^j]+[{\bf S}_2^i,{\bf S}_2^j] \right)
\,,
\end{eqnarray}
where the colour factor $C_A=N_c$, ${\bf q}={\bf p}-{\bf p}'$
and $\alpha_{{\tilde V}_{s}}$ contains the corrections to the static
potential up to NNLL~\cite{static,Pineda:2000gz}. We will set $c_4=1$ due to
reparameterization invariance. We would like to stress that the Wilson
coefficients are not dimensionless in $D\neq 4$.

The non-relativistic reduction of the spin operators of the potential
depends on the operators used to single out the physical state we want
to study. In our case we are using the vector currents, which project
to the spin-one state. We also have to be careful to use the same
conventions than those used to obtain the hard piece of the matching
coefficients. This produces $O(\epsilon)$ terms multiplying the ${\bf
S}^2$ operator. In practise one can do the following replacement for
the ${\bf S}^2$ operator with ${\bf S}^2=2$
\begin{equation}
- \frac{4\pi C_F D^{(2)}_{S^2,s}}{d\, m^2}\, 
       [{\bf S}_1^i,{\bf S}_1^j][{\bf S}_2^i,{\bf S}_2^j]
\to 
\frac{{\bf S}^2}{4}\frac{2\pi C_F D^{(2)}_{S^2,s}}{d\, m^2}\,
 \left[-(d-4)(d-1)\right] 
 \,.
\end{equation}
For S-wave creation there is no contribution from the potentials 
proportional to the $S_{12}$ and ${\bf L}\cdot{\bf S}$ operators.

The $D$-dimensional prescriptions used above are irrelevant for the
computation of the heavy quarkonium mass with NNLL accuracy, which
reflects the fact that the leading running of the Wilson coefficients
is scheme independent. On the other hand to {\it specify} the
$D$-dimensional prescription of the potential is important once they
are introduced into divergent potential loops. This is relevant if we
want to obtain heavy quarkonium sum rules, or to compute the $t
\bar{t}$ production near threshold with NNLL accuracy, and use
computations obtained in other places for the hard matching
coefficients.  We would like to emphasize however that we could have
chosen a different prescription. This would have changed some
intermediate-step results but not the physical results.

Finally, one should note that 
the potential above is not equal to the potential used in
Ref.~\cite{Beneke:1999qg}. Nevertheless, they can be exactly related
with each other by field redefinitions (in four and $D$
dimensions). In particular this means that the hard matching
coefficients will be the same in both cases.  This is actually what we
expected, since they simply correspond to the effects of the
(integrated out) hard modes.

The RGI coefficients of the various potentials are known to the
accuracy required for evaluating the Green function at
NNLL~\cite{Pineda:2000gz,Pineda:2001ra}.  Note that the strong
coupling is included in the matching coefficients. Thus $D^{(1)}_s
\simeq \alpha_s^2(1+(\alpha_s\log v)^n)$ and $D^{(2)}_X \simeq
\alpha_s(1+(\alpha_s\log v)^n)$. For the Coulomb potential
$\alpha_{{\tilde V}_{s}}$, the exact static potential at two
loop~\cite{static} receives additional three-loop LL terms
proportional to $\alpha_s^3 C_A^3 (\alpha_s \log
v)^n$~\cite{Pineda:2000gz}.  The explicit form of all matching
coefficients can be found in Ref.~\cite{Pineda:2001ra}.  However, we
have changed the basis of potentials compared to
Ref.~\cite{Pineda:2001ra}. This affects the matching coefficient
$D^{(2)}_{d,s}$, which now reads ($\mu_{us} =\mu_s^2/\mu_h$)
\begin{eqnarray}
D^{(2)}_{d,s}(\mu_{us}) &=& 
\frac{\alpha_s(\mu_s)}{3}\left(2+c_D(\mu_s)\right) 
+ \frac{1}{3\pi} \left( d_{vs}(\mu_s) + \frac{1}{C_F} d_{ss}(\mu_s) \right)
\nonumber
\\
&&
+\
\frac{32}{9\beta_0}\left( \frac{C_A}{2}-C_F \right) \alpha_s(\mu_s)
\log\left[\frac{\alpha_s(\mu_s)}{\alpha_s(\mu_{us})} \right]
\,.
\end{eqnarray}
For the other potentials, the expressions obtained for their Wilson
coefficients in Ref.~\cite{Pineda:2000gz,Pineda:2001ra} hold, where
one may also find the expressions for the RGI coefficients ($c_D$,
$d_{vs}$, $d_{ss}$, $\ldots$) of the NRQCD operators. We repeat them
in the Appendix for ease of reference.

In this paper we also include an additional contribution in
$\delta\tilde V$, the electromagnetic Coulomb term
\begin{equation}
\delta\tilde  V \to \delta\tilde  V 
- \frac{4\pi\, e_Q^2\, \alpha_{\rm EM}}{{\bf q}^2}
\,.
\label{EMpot}
\end{equation}
This will give rise to a NLO term $\sim \alpha_{\rm EM}/v$ from single
potential photon exchange and a NNLO term from double potential photon
exchange. Strictly speaking, the coupling in \Eqn{EMpot} should be
evaluated at the soft scale and not at the hard scale, but this effect
is beyond NNLL.

This completes all the corrections needed at the Lagrangian level for
the computation of the imaginary part of the Green function with NNLL
accuracy.  What is left is to obtain the RGI expressions for $c_s$ and
$d_s$, where $s=0$, 1 labels the spin.  The matching coefficient $c_s$
is needed at NNLL whereas $d_s$ is only needed at LL.

\subsection{Direct ultrasoft effects to $c_s$ and $d_s$}

Most of the ultrasoft contribution to the running of $c_s$ and $d_s$
comes in a indirect way, through the running of the potentials in the 
anomalous dimensions of the RG equation. Nevertheless, 
there are some genuine ultrasoft effects that have not been considered 
so far. They are due to the appearance of some energy dependent potentials 
with the structure
\begin{equation}
\label{deltaVus}
\delta V_{us} =(H_c-E)Z^{1/2}
\,,
\end{equation}
where $Z$ is the normalization correction to the 
heavy quarkonium propagator. Therefore, $Z^{1/2}$ 
corresponds to the normalization of the field that represents the 
heavy quarkonium. $\delta V_{us}$ was 
not included in Eq. (\ref{Hpnrqcd}), as there only energy independent
potentials were considered. 
Its effects could be reabsorbed in field
redefinitions of the fields that represent the heavy quarkonium.
Therefore, they have no consequences in the spectrum. Nevertheless,
these field redefinitions change the vertex interaction of the heavy
quarkonium with photons producing changes in $c_s$ and $d_s$ and they
have to be considered in our computation.  The leading logarithmic
corrections to $Z^{1/2}$ were obtained in Eqs.~(15--18) of
Ref.~\cite{Brambilla:1999xj} (for some partial results see also
Ref.~\cite{Kniehl:1999ud}).  They produce corrections of order
$\als^3\log\als$. We can obtain the RGI expressions for them and
generate terms of order $\als^{(n+3)}\log^{(1+n)}\als$. Note that the
running goes up to the soft scale because one gets expressions of the
form $\langle {\bf r}=0|\log r|{\bf p}\rangle$ and the logarithm gets
the scale of $p$. The corrections to the Green function due to
\Eqn{deltaVus} read
\begin{eqnarray}
\delta G_1&=&{4 \als^2 \over \beta_0}
\log{\left[\als(\mu_s)\over \als(\mu_{us})\right]}
{1 \over H_c-E}{C_A^2C_F \over 4}\,,
\\
\delta G_2&=&{4 \als \over \beta_0}
\log{\left[\als(\mu_s)\over \als(\mu_{us})\right]}
\left\{{1 \over mr},{1 \over H_c-E}\right\}
\left({2 \over 3}C_F^2+C_FC_A\right)
\,,
\label{G2}\\
\label{deltaG3}
\delta G_3&=&{4 \over \beta_0}
\log{\left[\als(\mu_s)\over \als(\mu_{us})\right]}
\left\{{{\bf p}^2 \over m^2},{1 \over H_c-E}\right\}{2 \over 3}C_F
\,.
\end{eqnarray}
The correction coming from $\delta G_3$ can be included in $d_s$:
\be
\label{USds}
d_s \rightarrow d_s+ \frac{16 C_F}{\beta_0} \, 
\log\left[{\als(\mu_{us})\over \als(\mu_s)}\right]
\,.
\ee
The
correction from $\delta G_2$ is zero. Finally, the correction from
$\delta G_1$ could be absorbed in $c_s$:
\begin{equation}
c_s \rightarrow c_s+\frac{\als^2}{2}\frac{C_A^2C_F}{\beta_0}
\log{\left[\als(\mu_s)\over \als(\mu_{us})\right]}
\,.
\end{equation}
Note also that these changes are equivalent to a change in the NNLO
and LO anomalous dimension of the RG equation describing the running
of $c_s$ and $d_s$ respectively.

\subsection{Running of $d_s$}

The LL running of ${\rm Im} g_{\rm EM}^{\rm pNR}$ can be obtained in
two steps. In the first step one computes its soft running. This has
been done in Ref.~\cite{Bodwin:1994jh,Brambilla:2002nu}.  For the
explicit result see Eqs.~(C.18,C.19) in Ref.~\cite{Brambilla:2002nu}.
The ultrasoft running can be obtained from \Eqn{USds}.  Adding
everything together the LL running of ${\rm Im} g_{\rm EM}^{\rm
pNR}$ reads
\begin{eqnarray}
\lefteqn{{\rm Im\,}g^{\rm pNR}_{\rm EM}(^1S_0)(\mu_s) = } && \\
&& \nonumber
{\rm Im\,}g^{\rm pNR}_{\rm EM}(^1S_0)(\mu_h)-
{16 \over 3\beta_0}C_F\, 
{\rm Im\,} f^{\rm pNR}_{\rm EM}(^1S_0)(\mu_h)
\log\left[{\als(\mu_{us})\over \als(\mu_h)}\right] 
\,,
\\
\lefteqn{{\rm Im\,}g^{\rm pNR}_{\rm EM}(^3S_1)(\mu_s) = } && \\
&& \nonumber
{\rm Im\,}g^{\rm pNR}_{\rm EM}(^3S_1)(\mu_h)-{16 \over 3\beta_0}C_F\, 
{\rm Im\,}f^{\rm pNR}_{\rm EM}(^3S_1)(\mu_h)
\log\left[{\als(\mu_{us})\over \als(\mu_h)}\right] 
\end{eqnarray}
The last result agrees with the LL result obtained in
Ref.~\cite{Hoang:2001mm}. For the matching coefficient $d_1$ as
defined in \Eqn{NRQCDcurrent}, this entails
\begin{equation}
\label{d1run}
d_1(\mu_{us}) = 1 + \frac{16 C_F}{\beta_0} \, 
\log\left[{\als(\mu_{us})\over \als(\mu_h)}\right]
\,.
\end{equation}

\subsection{Running of $c_s$}

The running of $c_s$ is not yet known
with NNLL accuracy. It is dictated by the solution of the RG equation
(the LO anomalous dimension is zero)
\begin{equation}
\mu_s \frac{d}{d \mu_s}\log c_s
=
\gamma_{c_s}^{\rm NLO}
+
\gamma_{c_s}^{\rm NNLO}
+
\cdots
.
\end{equation}
The structure of the solution reads
\begin{equation}
\label{cs}
c_s(\mu_s)
= c_s(\mu_h)
  e^{\alpha_s(\mu_h)\Gamma_{ c_s}^{\rm NLL}(\mu_s)
    +\alpha_s^2(\mu_h)\Gamma_{ c_s}^{\rm NNLL}(\mu_s)+\cdots}
  \,.
\label{csnnl}
\end{equation}
Expressions for $c_s(\mu_h)$ at two loops in the $\MS$ 
can be found in Ref. \cite{c1nnlo} for $s=1$ and (almost complete) 
in Ref. \cite{Czarnecki:2001zc} for $s=0$. 

The expression for $\gamma_{c_s}^{\rm NLO}$ in the basis of potentials 
used in this paper reads
\begin{equation}
\gamma_{c_s}^{\rm NLO}=-\frac{C_F^2}{4}\als
\left[
\als+\left(2-\frac{4}{3}s(s+1)\right)
D_{S^2,s}^{(2)}-3D_{d,s}^{(2)}+4D_{1,s}^{(2)}\right]
-\frac{C_A C_F}{2}D_{s}^{(1)}
\,.
\end{equation}
The expression of $\Gamma_{ c_s}^{\rm NLL}$ is known
\cite{Pineda:2001et,Hoang:2002yy}.  For $\Gamma_{c_s}^{\rm NNLL}$ only
the spin-dependent term is completely known \cite{Penin:2004ay}.  Its
contribution to the spin-one case reads
\begin{equation}
\label{SD}
\delta \Gamma_{c_1,SD}^{\rm NNLL}(\mu_s)=\Gamma_{\hat c_v}^{\rm NNLL}
+{11 \over 72}C_F^2\, {\als(\mu_s) \over \als^2(\mu_h)} 
D^{(2)}_{S^2,s}(\mu_s)
-{11 \over 72}C_F^2
\,,
\end{equation}
where $\Gamma_{\hat c_v}^{\rm NNLL}$ corresponds to the result 
quoted in Ref. \cite{Penin:2004ay}.
Note that we use a different expression for the spin-dependent term than 
just $\Gamma_{\hat c_v}^{\rm NNLL}$. The expression above corresponds 
to the spin-dependent contribution to the vector current matching 
coefficient in the $\MS$ scheme. Numerically this contribution is 
small compared with others.

Besides the spin-dependent correction, we have also incorporated the 
following spin-independent corrections at NNLL order (to these 
corrections one obviously has to subtract the spin-dependent piece that 
has already been included in $\delta \Gamma_{c_1,SD}^{\rm NNLL}$):\\ 
a) those that appear from the exponentiation of the NLL term and
  formally are NNLL (see \Eqn{cs}),\\
b) Effects due to the two-loop 
beta running of $\als$. They produce the following correction:
\begin{eqnarray}
\als^2(\mu_h)\delta \Gamma_{c_1,b}^{\rm NNLL}(\mu_s)&=&
\frac{\beta_1}{2\beta_0^2}
\int_{\als(\mu_h)}^{\als(\mu_s)}\frac{d\als}{\als}\gamma_{c_s}^{\rm NLO}
\\
\nn
&&
-\frac{2\pi}{\beta_0}\int_{\als(\mu_h)}^{\als(\mu_s)}\frac{d\als}{\als}
\beta_1\frac{\partial \als(\mu_s^2/\mu_h)}{\partial \beta_1}
\frac{\partial \gamma_{c_s}^{\rm NLO}}{\partial \als(\mu_s^2/\mu_h)}
\,.
\end{eqnarray}
The last term is generated from the fact that in the determination of 
$\Gamma_{ c_s}^{\rm NLL}(\mu_s)$, the relation 
\begin{equation}
\frac{\als(\mu_s^2/\mu_h)}{\als(\mu_s)} \rightarrow 
\frac{1}{(2-z^{\beta_0})}
\,,
\end{equation}
where $z^{\beta_0}=\als(\mu_s)/\als(\mu_h)$, was used. This relation is
only true at one loop and has to be corrected by $\beta_1$ terms if a
NNLL accuracy is demanded. The numerical impact of these corrections
is small.\\ 
c) We have also incorporated the corrections
proportional to $a_1$, the one-loop log-independent term, that appears
in $\al_{V_s} \simeq
\als(\mu_s)\left(1+\frac{\als}{4\pi}a_1\right)$. These corrections can
be deduced from the computation of the NLO anomalous dimension.  They
read
\begin{equation}
\als^2(\mu_h)\delta \Gamma_{c_1,c}^{\rm NNLL}(\mu_s)
=
-\frac{1}{2\beta_0}a_1\int_{\als(\mu_h)}^{\als(\mu_s)}\frac{d\als}{\als}
\left(
\gamma_{c_s}^{\rm NLO}+\frac{C_AC_F}{2}D_s^{(1)}-C_F^2\frac{\als^2}{4}
\right)
\,.
\end{equation}
The numerical impact of these corrections is small.

Finally, the inclusion of the electromagnetic corrections produces
some corrections to $c_1$.  Counting $\alpha_{\rm EM}\sim \alpha_s^2$,
the one-loop exchange of a hard photon contributes at NNLO and is
taken into account by
\begin{equation}
\label{c1qed}
c_1(\mu_h) \to c_1(\mu_h) - \frac{2\, e_Q^2 \alpha_{\rm EM}}{\pi}
\end{equation}
for the spin-one case, whereas for $s=0$ we have
\begin{equation}
\label{c0qed}
c_0(\mu_h) \to c_0(\mu_h) - \frac{e_Q^2 \alpha_{\rm EM}}{\pi}
\left(\frac{5}{2} - \frac{\pi^2}{8} \right).
\end{equation}

\subsection{Green function}

Once the RGI potential and current matching coefficients are
available, we are in a position to use standard quantum mechanics
perturbation theory to compute the higher-order corrections to ${\rm
Im}[G({\bf 0}, {\bf 0};E)]$ via insertions of the potentials. This
calculation has been done in momentum space using dimensional
regularization. For the terms suppressed by two powers of
$\alpha_s\sim v$ in $\delta \tilde{V}$, Eq.(\ref{Hpnrqcd}), it is
sufficient to consider a single insertion,
\begin{equation}
\label{singleIns}
\delta G({\bf 0},{\bf 0};E) = \int \prod_{i=1}^4 \frac{d^d{\bf p}_i}{(2\pi)^d}
\, \tilde{G}_c({\bf p}_1,{\bf p}_2;E)\, 
\delta\tilde{V}({\bf p}_2,{\bf p}_3)\, 
\tilde{G}_c({\bf p}_3,{\bf p}_4;E)
\,,
\end{equation}
whereas for terms suppressed by only a single power of $\alpha_s\sim
v$ we have to compute double insertions as well,
\begin{eqnarray}
\label{doubleIns}
\lefteqn{\delta G({\bf 0},{\bf 0};E) = } & & \\
& & \int \prod_{i=1}^6 \frac{d^d{\bf p}_i}{(2\pi)^d}
\, \tilde{G}_c({\bf p}_1,{\bf p}_2;E)\, 
\delta\tilde{V}({\bf p}_2,{\bf p}_3)\, 
\tilde{G}_c({\bf p}_3,{\bf p}_4;E)\,
\delta\tilde{V}({\bf p}_4,{\bf p}_5)\, 
\tilde{G}_c({\bf p}_5,{\bf p}_6;E)
\,.
\nonumber
\end{eqnarray}
Since the RGI does not alter the
structure of the interaction terms, the insertions can be taken
directly from Ref.~\cite{Beneke:1999qg}. In Appendix~\ref{AppIns} we
list the integrals that are needed and present the explicit results
obtained in Ref.~\cite{Beneke:1999qg}.

In the calculation described so far, the on-shell scheme for the heavy
quark mass has been implicitly assumed.  In order to avoid the bad
convergence behaviour inherent to this scheme, it is necessary to
rewrite the expressions in terms of a threshold mass which is free of
the renormalon ambiguity. In this paper we will consider the cases of
the potential subtracted (PS) mass~\cite{Beneke:1998rk} and renormalon
subtracted (RS) mass~\cite{Pineda:2001zq}. We will use the difference
between both schemes as an indication of the scheme dependence of our
results.

\section{The case of the top quark}

In this section we apply the previous results to the case of top-quark
pair production near threshold at a future linear collider. Due to the
large width $\Gamma_t$ of the top quark there are no bound states and
the toponium resonances are smeared, resulting in a smooth curve for
the cross section with a broad peak as the remnant of the would-be
$1S$ bound state. Using perturbation theory we can reliably compute
the cross section as a function of the energy~\cite{FadinKhoze}. 
This may lead to accurate determinations of the top mass and its 
total decay width by measuring the position and normalization of the 
peak.

The $t\bar{t}$ pair will be dominantly produced via $e^+ e^- \rightarrow
\gamma^\ast \, ,\, Z^\ast \rightarrow t\bar t$.  The total production cross
section may be written as \cite{Hoang:2000yr}
\begin{eqnarray}
  \sigma_{\rm tot}^{\gamma,Z}(s) = \frac{4\pi\alpha_{\rm EM}^2}{3 s} \Big[\,
  F^v(s)\,R^v(s) + F^a(s) R^a(s) \Big] \,,
\label{totalcross}
\end{eqnarray}
where
\begin{eqnarray} \label{fullR}
 R^v(s) \, = \,\frac{4 \pi }{s}\,\mbox{Im}\,\left(-i\int d^4x\, e^{i q\cdot x}
  \langle\,0 | T\{ j^v_{\mu}(x) \, {j^v}^{\mu}(0)\}| 0\rangle \right) \,, 
\nonumber\\
 R^a(s) \, = \,\frac{4 \pi }{s}\,\mbox{Im}\,\left(-i\int d^4x\, e^{i q\cdot x}
  \langle\,0 | T\{ j^a_{\mu}(x) \, {j^a}^{\mu}(0)\}| 0\rangle \right) \,, 
\end{eqnarray}
and $j^{v}_\mu$ ($j^{a}_\mu$) is the vector (axial-vector) current
that produces a quark-antiquark pair defined by
Eq.~(\ref{NRQCDcurrent}). With both $\gamma$ and $Z$ exchange the
prefactors in Eq.~(\ref{totalcross}) are
\begin{eqnarray}
  F^v(s) &=& \bigg[\, e_q^2 - 
   \frac{2 s\, v_e v_q e_q}{s-m_Z^2} + 
   \frac{s^2 (v_e^2+a_e^2)v_q^2}{(s-m_Z^2)^2}\, \bigg]\,,
\nonumber\\[2mm]
  F^a(s) &=& \frac{s^2\, (v_e^2+a_e^2)a_q^2}{ (s-m_Z^2)^2 } \,,
\end{eqnarray}
where 
\begin{equation}
  v_f = \frac{T_3^f-2 e_f \sin^2\theta_W}{2\sin\theta_W \cos\theta_W}\,,
  \qquad\qquad
  a_f = \frac{T_3^f}{2\sin\theta_W \cos\theta_W} \,.
\end{equation}
Here $e_f$ is the charge for fermion $f$, $T_3^f$ is the third
component of weak isospin, $\theta_W$ is the weak mixing angle, and
$m_Z$ the mass of the $Z$.  Here we will focus on $R^v$, since it
gives the dominant contribution and we are mainly interested in
studying the impact of the resummation of logarithms on the
convergence and scale dependence of the perturbative expansion of
$R^v$. A study of $R^a$ can be found in Ref. \cite{Hoang:2001mm}.

Since $\Gamma_t\sim m \alpha_{\rm EM} \sim m v^2$ and the propagator of a
potential heavy quark scales as $v^{-2}$ the effects due to the width
of the top quark are LO effects and have to be taken into
account by modifying the propagator $E-{\bf p}^2/(2m) \to
E+i\Gamma_t - {\bf p}^2/(2m)$. This amounts to replacing $E\to
E+i\Gamma_t$ in \Eqn{L0pnrqcd}.  As noted in Ref.~\cite{FadinKhoze},
this is only correct at LO. Higher-order electroweak
corrections have a much richer structure and it is not possible any
longer to formulate the problem in terms of a $t\bar{t}$ final
state. In particular, at NNLO there are interference effects (between
double and single resonant processes), QED radiation effects and
non-factorizable corrections (non-trivial interconnections between the
decay products of the top quarks with the remainder of the
process). Any consistent approach beyond LO has to
introduce additional operators with fields corresponding to the
incoming electrons and the decay products of the top quarks, and link
higher-order corrections of the $\psi^\dagger \psi$ and $\chi^\dagger
\chi$ operators to three-point and higher-point vertices. Even though
there is an effective theory framework available for systematically
taking into account these corrections~\cite{upet}, a full explicit
calculation of electroweak effects at NNLO is still lacking. If we are
interested in the total cross section only, the situation is somewhat
simpler, since the non-factorizable corrections cancel~\cite{nfc}, 
and the replacement $E\to E+i\Gamma_t$ becomes correct with NLO accuracy.
Furthermore, some of the electroweak corrections have be taken into
account by including them in the matching
coefficients~\cite{Hoang:2004tg}.  In this article we restrict
ourselves to the usual shift $E\to E+i\Gamma_t$.

\begin{figure}[h!]
   \begin{center}
   \includegraphics[width=.75\textwidth]{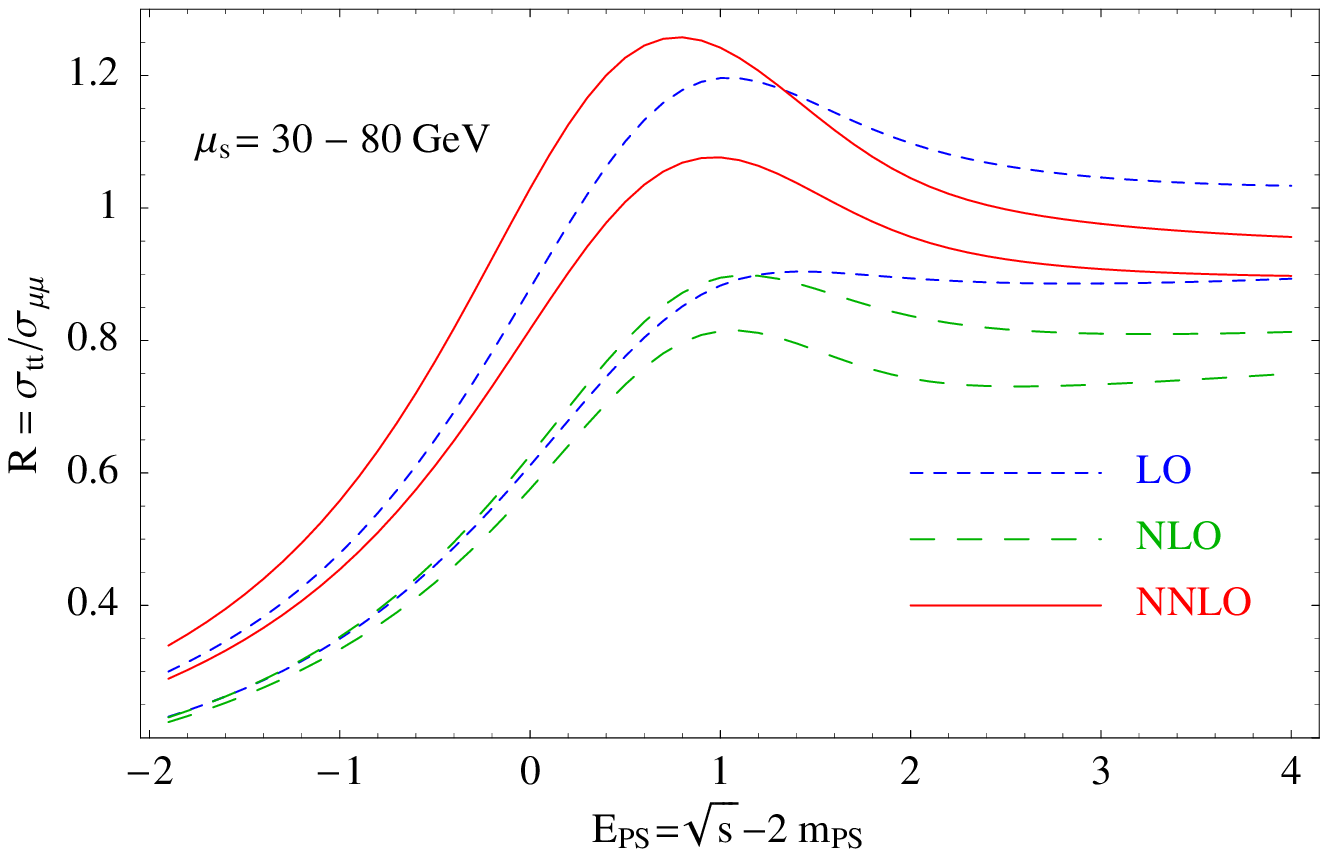} \\[10pt]
   \includegraphics[width=.75\textwidth]{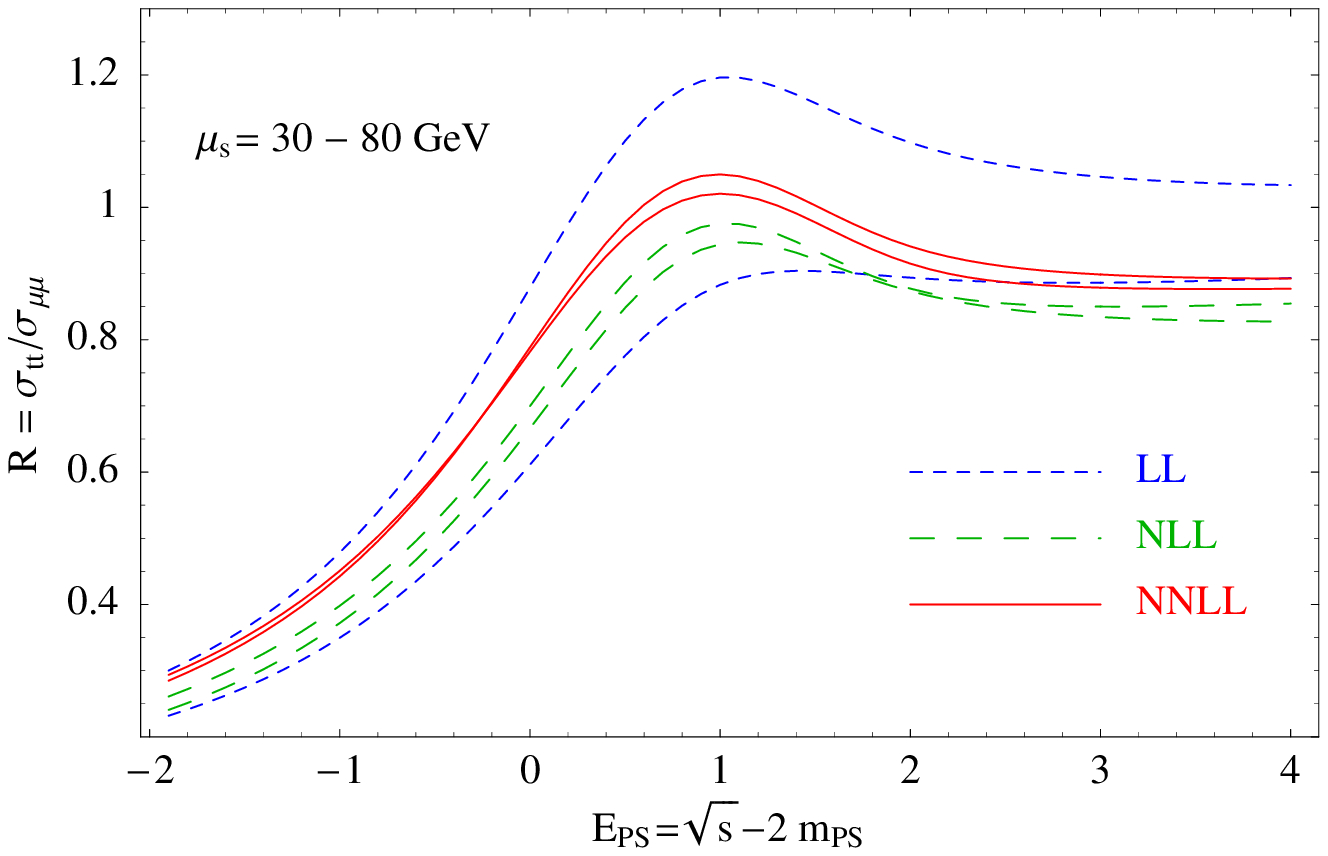}
   \caption{\small{
   Threshold scan for $t\bar{t}$ using the PS mass, $m_{\PS}(20\,
   {\rm GeV})=175$ GeV. The upper
   panel shows the fixed order results, LO, NLO and NNLO, whereas in
   the lower panel the RGI results LL, NLL and NNLL are displayed. The
   soft scale is varied from $\mu_s$=30~GeV to $\mu_s$=80~GeV. }}
   \label{fig:scan1}
   \end{center}
\end{figure}

In Figure~\ref{fig:scan1} the fixed-order results are compared to the
RGI results obtained using the procedure described above and using the
PS mass with the subtraction scale $\mu_F$ required for the definition
of the threshold mass set to 20~GeV. The plots were produced with an
energy dependent soft scale $\mu_s^2 = 4 m \sqrt{E^2+\Gamma_t^2}$. In
order to obtain a first rough estimate of the theoretical uncertainty
we show the cross sections as bands obtained by variation of the soft
scale in the region $30~{\rm GeV} \le \mu_s \le 80~{\rm GeV}$, where
these numbers refer to the scale $\mu_s$ at $E=0$. For the hard and
ultrasoft scale we take our default values $\mu_h = m_{\PS} = 175$~GeV
and $\mu_{us} = \mu_s^2/\mu_h$. Our results are qualitative
consistent, but not equal, to those obtained in
Refs.~\cite{Hoang:2000ib, Hoang:2001mm}.  This agreement is not
trivial, since the ingredients included in the NNLL analysis are
different in ours and their computations. This may indicate that, even
if the NNLL evaluation is incomplete, the qualitative features will
hold in the complete result. We observe that the scale dependence is
much reduced once the logarithms are taken into account and reduces
from LL to NLL to NNLL.  We also note that the size of the corrections
decreases for the RGI results, in particular the NNLL band is much
closer to the NLL band.  However, the NNLL band does not overlap with
the NLL band, indicating that a theoretical error estimate relying on
the scale dependence alone is too optimistic. Therefore, we consider
other possible source of errors in what follows.

\begin{figure}[ht!]
   \begin{center}
   \includegraphics[width=.75\textwidth]{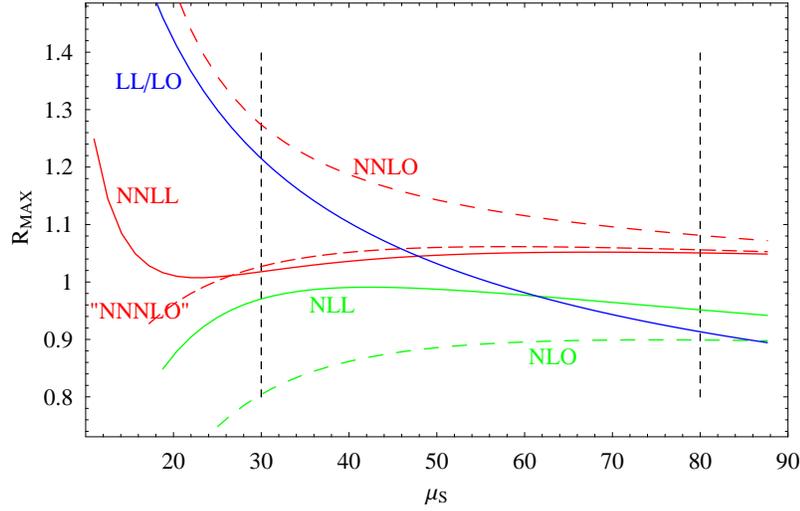} \\
   \caption{\small{The normalization of the peak of the RGI threshold
   cross section as a function of the soft scale $\mu_s$. The vertical
   dashed lines show the limits of variation used in
   Figure~\ref{fig:scan1}.} }
   \label{fig:peakmuS}
   \end{center}
\end{figure}

The variation of the soft scale has been stopped at $\mu_s=30$~GeV
which might seem to be a rather large value. In fact, the dependence
of the normalization of the peak as a function of $\mu_s$ is very
smooth at NNLL, up to a value of $\mu_s \sim 25$~GeV, where it
abruptly changes\footnote{As we will see a similar pattern also
appears for the bottomonium decays.}. At NLL, already for $\mu_s \sim
30$~GeV there is a rather large variation of the normalization. This
is illustrated in Figure~\ref{fig:peakmuS} which shows the $\mu_s$
dependence of the normalization of the peak at LO/LL, NLO, NLL, NNLO
and NNLL. As observed in Ref.~\cite{Beneke:2005hg} the situation for
small soft scales may be remedied if multiple insertions of the
Coulomb potential are taken into account. Since in our result these,
formally, higher-order contributions are not taken into account, we
refrain from using scales $\mu_s \le 30$~GeV and take $\mu_s=40$~GeV
as our default value for the soft scale unless stated otherwise.  The
``NNNLO'' result also depicted in Figure~\ref{fig:peakmuS} is obtained
by re-expanding the full RGI result and keeping only terms that are
NNNLO, but dropping those of even higher order. The difference between
this result and the full RGI result is small except for scales
$\mu_s \le 30$~GeV, which is outside the range we use.

\begin{figure}[h!]
   \medskip
   \begin{center}
   \includegraphics[width=.75\textwidth]{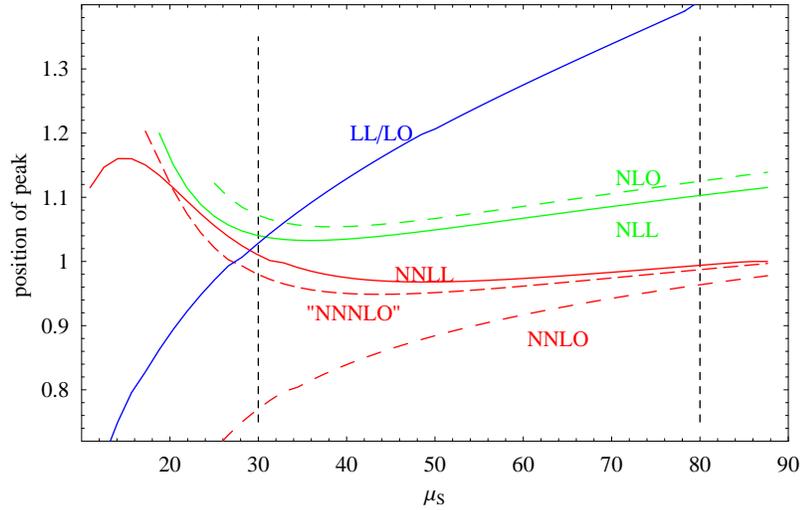} \\
   \caption{\small{The position of the peak of the RGI threshold cross
   section as a function of the soft scale $\mu_s$. The vertical
   dashed lines show the limits of variation used in
   Figure~\ref{fig:scan1}.}}
   \label{fig:pospeakmuS}
   \end{center}
\end{figure}

We also present a similar plot to show the scale dependence of the
position of the peak in Figure~\ref{fig:pospeakmuS}. From this plot we
can see that the NNLL resummation of logarithms significantly improves
over the fixed-order NNLO evaluation. The scale dependence is reduced
and also the convergence of the perturbative series is better, which
is reflected in a smaller difference between the NLL and NNLL result
versus the NLO and NNLO result (this also happens for the
normalization of the total cross section). The difference between the
``NNNLO'' and NNLL curve is again very small for $\mu_s\ge 30$~GeV.
From this plot we estimate the theoretical error for the determination
of the position of the peak (which is related to the determination of
the top mass) to be of the order of 100~MeV.

\begin{figure}[h!]
   \medskip
   \begin{center}
   \includegraphics[width=.75\textwidth]{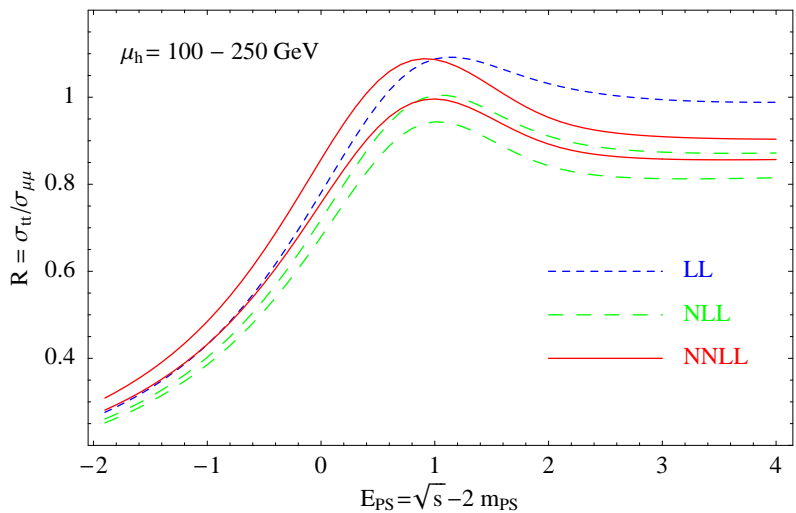}
   \caption{\small{Dependence of the $t\bar{t}$ threshold scan on the
   hard scale $\mu_h$, using the PS mass. At NNLL (NLL) the lower
   (upper) curve corresponds to $\mu_h=250$~GeV, whereas the upper
   (lower) curve corresponds to $\mu_h=100$~GeV.}}
   \label{fig:scan2}
   \end{center}
\end{figure}

In the fixed-order calculations the error is usually estimated from
the soft-scale dependence, even though one may think that the dominant
uncertainty came from the magnitude of the correction. This is
potentially more of a problem after the resummation of logarithms,
since the soft scale dependence is much less severe. Therefore, we
have to be more careful with other sources of uncertainties. In
particular, the missing ultrasoft contributions make it important to
consider the dependence on the other scales as well. Since they are
correlated, we can only vary $\mu_h$ together with $\mu_{us}$ and,
thus, consider in Figure~\ref{fig:scan2} the dependence on the hard
scale $\mu_h$, setting $\mu_s= 40$~GeV. Variation of the hard scale
around its natural value $\mu_h=m$ by choosing $100~{\rm GeV} \le
\mu_h \le 250~{\rm GeV}$ results in a scale dependence that is
considerably larger than the soft scale dependence. We note that this
error is compatible with the magnitude of the difference between the
NLL and NNLL result. It is to be expected that the situation improves
once all ultrasoft logarithms at NNLL are taken into account, but at
this stage the rather large dependence of the cross section on $\mu_h$
has to be taken into account if a theoretical error is assigned. We
also note that the dependence on $\mu_h$ only enters at NLL, thus the
LL ``band'' in Figure~\ref{fig:scan2} is simply a line.

\begin{figure}[h!]
   \medskip
   \begin{center}
    \includegraphics[width=.75\textwidth]{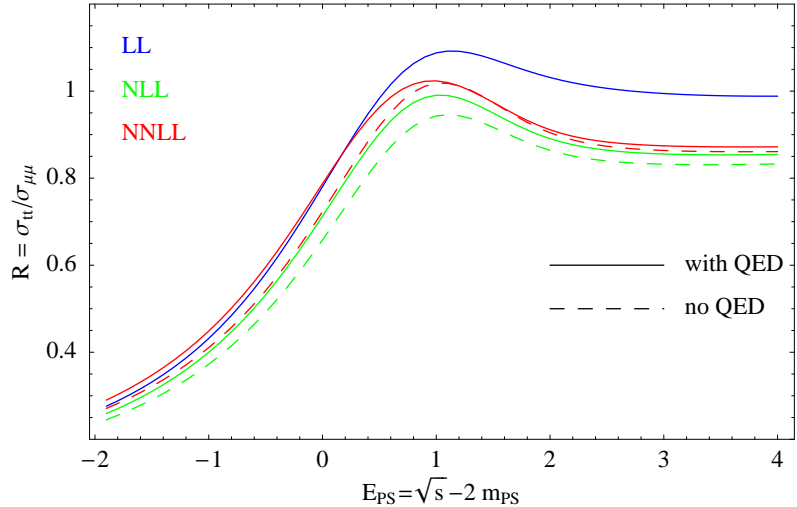}
   \caption{\small{Effects of the QED corrections to the $t\bar{t}$
   threshold scan. The hard ans soft sales are chosen as
   $\mu_h=m_{\rm PS}=175$~GeV and $\mu_s=40$~GeV.}}
   \label{fig:scan3}
   \end{center}
\end{figure}

Finally we turn to Figure~\ref{fig:scan3}, where we display the
effects due to the QED corrections with the default choice for the
scales, $\mu_h=m$ and $\mu_s=40$~GeV. As previously mentioned, the QED
effects enter at NLL (thus there is no effect on the LL curve) and,
compared to the desired accuracy (top quark mass measurement with an
error $\delta m \le 100$~MeV), they are large. They change the
normalization by up to 10\% and result in a shift in the extracted
$\overline{\rm MS}$ mass of up to 100~MeV, making it mandatory to
include them. We note that we have not changed the definition of the
PS mass. Thus the shifts shown in Figure~\ref{fig:scan3} are physical
effects and are not compensated if the PS mass is related to the
$\overline{\rm MS}$ mass.

\begin{figure}[h!]
   \medskip
   \begin{center}
   \includegraphics[width=.75\textwidth]{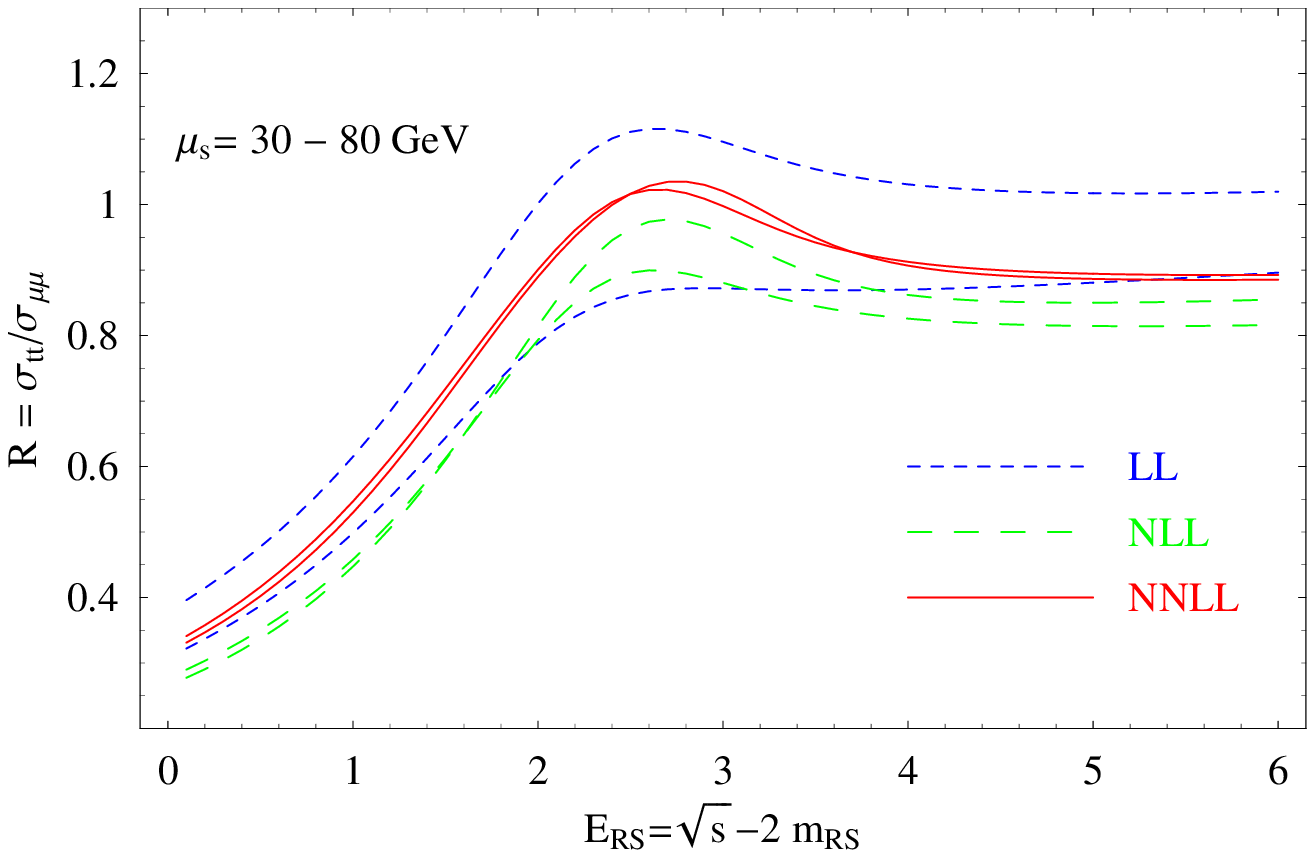} \\[10pt]
   \includegraphics[width=.75\textwidth]{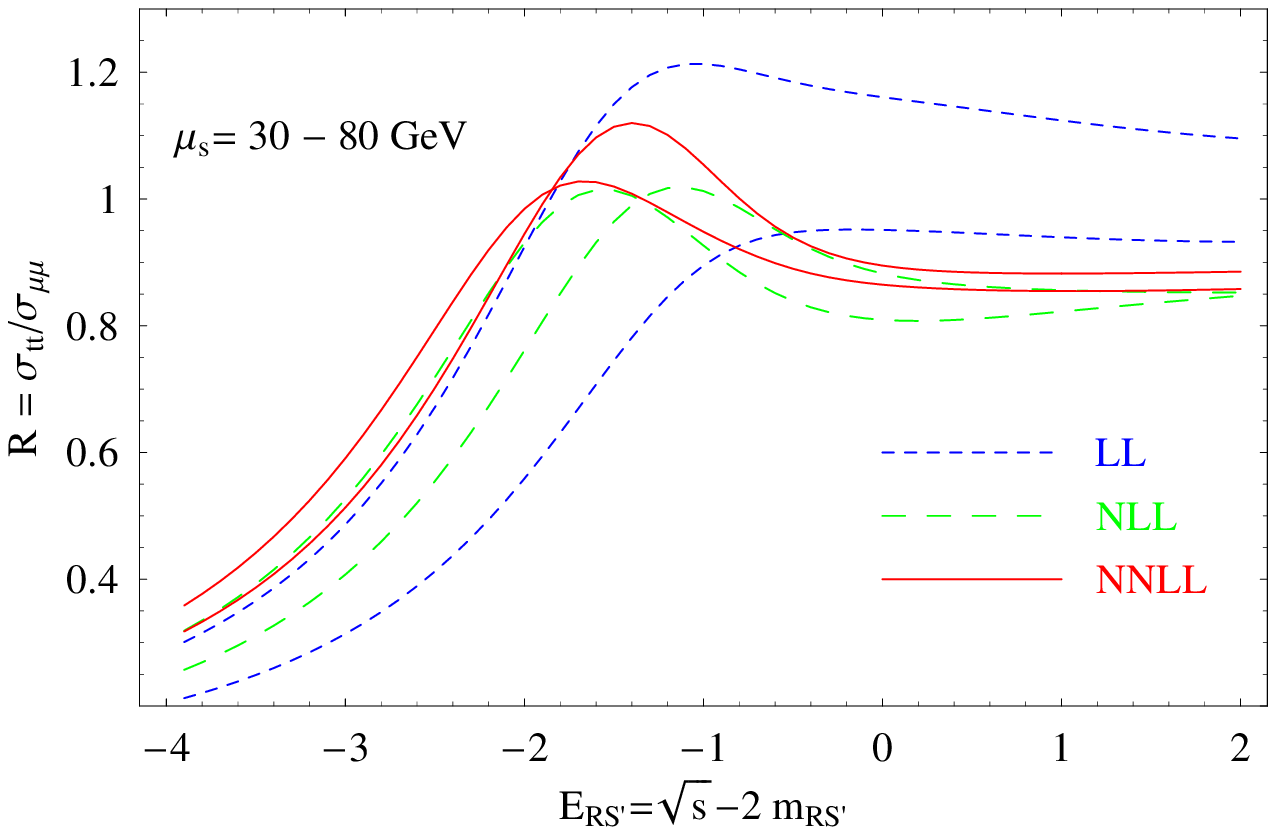}
   \caption{\small{Threshold scan for $t\bar{t}$ using the RS and RS'
   mass, $m_{\rm RS}(20\, {\rm GeV})=175$~GeV, $m_{\rm RS'}(20\, {\rm
   GeV})=175$~GeV. The upper panel shows the RGI results LL, NLL and
   NNLL with the RS and the lower panel with the RS' mass. The soft
   scale is varied from $\mu_s$=30~GeV to $\mu_s$=80~GeV. }}
   \label{threshold}
   \end{center}
\end{figure}

We end this section by mentioning that all results presented here can
be worked out in different threshold mass schemes. The qualitative
features of the results obtained are similar. We illustrate this point
using the RS and RS' mass, again setting $\mu_F=20$~GeV, and depict
the threshold scan in Figure~\ref{threshold}. Those plots, together
with the plot in the PS scheme (lower panel in
Figure~\ref{fig:scan1}), allow us to visualize the effect of using
different threshold masses. In principle, the value of the threshold
masses can run from being numerically close to the $\MS$ mass (but not
too close otherwise power counting is broken) to being numerically
close to the pole mass (but again not too close, otherwise the
renormalon cancellation does not take place). Within this allowed
range, the threshold mass definitions numerically closer to the $\MS$
mass have a smaller scale dependence. The price paid is that the
magnitude of the corrections (between different order in perturbation
theory) is larger. For the specific threshold masses we use, the RS
mass is the one closest to the $\MS$ mass whereas the RS' mass is
closest to the pole mass. The PS mass is in the middle, though
somewhat closer to the RS mass.  Once experimental data is available
it will be useful to consider different mass definitions to obtain an
estimate of the corresponding error.

\section{The case of the bottom quark}

We can also apply the results of Section~2 to a variety of observables
for $b\bar b$ systems. Likely, the theoretically cleanest observable
on which one can use our results is non-relativistic sum rules, which
has already been considered in Ref.~\cite{Pineda:2006gx}, where an
accurate determination of the bottom mass was obtained. Here we will
consider inclusive electromagnetic decays (either to $e^+e^-$ or to
$\gamma\gamma$) of the bottomonium ground state for which we will
provide analytic formulae.  These can be obtained from the results
obtained for the non-relativistic Green function. The spin-one decay
has the following structure
\begin{eqnarray}
\label{vector}
\lefteqn{\Gamma(\Upsilon(nS) \rightarrow e^+e^-) = } && \\
\nonumber &&
16\pi \, 
{C_A \over 3} \left[ { \alpha_{EM}\, e_Q \over M_{\Upsilon(nS)}} \right]^2
\left|\phi^{(s=1)}_n({\bf 0}) \right|^2
\left\{ c_1-d_1{M_{\Upsilon(nS)}-2m \over 6m} \right\}^2
\,.
\end{eqnarray}

The corrections to the wave function at the origin are obtained by
taking the residue of the Green function at the position of the poles
\begin{equation}
\left|\phi^{(s=1)}_n({\bf 0}) \right|^2
=
\left|\phi^{(0)}_n({\bf 0})\right|^2
\left(1+\delta \phi^{(s=1)}_n\right)
=
{\
\vbox{\hbox{\rlap{Res}\lower9pt\vbox
{\hbox{$\scriptscriptstyle{E=E_n}$}}}}\
} \! 
 G_{s=1}({\bf 0},{\bf 0};E)
\,,
\end{equation}
where the LO wave function is given by
\begin{equation}
\left|\phi^{(0)}_n({\bf 0})\right|^2=
{1 \over \pi} \left({mC_F\als \over 2n }\right)^3.
\end{equation}
The corrections to $\delta \phi_n^{(s=1)}$ 
produced by $\delta V$ have already been calculated with NNLO 
accuracy \cite{Melnikov2,Penin} in the direct matching scheme. One can 
also obtain them in the dimensional regularized $\MS$ scheme with 
NNLL accuracy by incorporating the RGI matching coefficients. One obtains 
then the following correction to the wave function 
\beqa
&&
\delta \phi_n^{(s=1)}
= \frac{\als^2C_A^3}{2\beta_0}
\log{\left[\als(\mu_s)\over \als(\mu_{us})\right]}
\\
\nn
&&
+ \, {\als \over \pi}
\left(\frac{3\, a_1}{4} +
{\beta_0\over 2}
\left(3L[n]
    + S[1,n] + 2 n S[2,n]-1-\frac{n \pi^2}{3}\right) \right)
\\
\nn
&&
+ \, C_FC_A D_s^{(1)}\,
\left(L[n]
- S[1,n] + {2\over n}+{5 \over 4} \right)
\\
\nn
&&
+ \, 2 C_F^2 \alpha_s  D_{1,s}^{(2)}
\left(L[n] - S[1,n] 
- \frac{5}{8n^2} + \frac{2}{n} + \frac{3}{2}  \right)
\\
\nn
&&
- \, {C_F^2 \alpha_s  D_{S^2,s}^{(2)} \over 3}
\left(L[n] - S[1,n]
+{2\over n} + {11 \over 12} \right)
\\
\nn
&&
- \, {3C_F^2 \alpha_s  D_{d,s}^{(2)} \over 2}
\left(L[n] - S[1,n]
+{2\over n}+{1 \over 2} \right)
\\
\nn
&&
- \, {C_F^2 \over 4}D_{2,s}^{(2)}\als
\\
\nn
&&
+ \, c_4{C_F^2  \als^2 \over 2}
\left(L[n] - S[1,n]
- \frac{3}{4n^2} + \frac{2}{n} + \frac{3}{2} \right)
\\
\nn
&&
+\frac{\als^2}{(4\pi)^2}
\bigg(
3a_1^2 + 3a_2 - 14a_1\beta_0 + 4\beta_0^2 - 2\beta_1 + \beta_0^2\pi^2 
\\
\nn
&&
- \frac{8a_1\beta_0n\pi^2}{3} + \frac{4\beta_0^2n\pi^2}{3} 
- \frac{2\beta_1n\pi^2}{3} 
+ \frac{\beta_0^2n^2\pi^4}{9} + 24a_1\beta_0
  L[n] 
\\
\nn
&&
- 28\beta_0^2L[n] 
+ 6\beta_1L[n] 
- \frac{16\beta_0^2n\pi^2}{3}L[n] 
+ 24\beta_0^2L[n]^2 
\\
\nn
&&
+ 8a_1\beta_0S[1, n] - 20\beta_0^2S[1, n] + 2\beta_1S[1, n] 
- \frac{12\beta_0^2S[1, n]}{n} - \frac{8\beta_0^2n\pi^2S[1, n]}{3} 
\\
\nn
&&
+ 16\beta_0^2L[n]S[1, n] 
+ 8\beta_0^2S[1, n]^2 
+ 8\beta_0^2S[2, n] + 16a_1\beta_0nS[2, n] 
\\
\nn
&&
- 8\beta_0^2 n S[2, n] 
+ 4\beta_1 n S[2, n] - \frac{4\beta_0^2n^2\pi^2S[2, n]}{3} 
+ 32\beta_0^2nL[n]S[2, n] 
\\
\nn
&&
+ 
   16\beta_0^2nS[1, n]S[2, n] + 4\beta_0^2n^2S[2, n]^2 + 28\beta_0^2nS[3, n] 
   - 20\beta_0^2n^2S[4, n] 
   \\
\nn
&&
- 24\beta_0^2nS_2[2, 1, n] + 16\beta_0^2n^2S_2[3, 1, n] + 
   20\beta_0^2n\, \zeta(3)\bigg)
\,,
\end{eqnarray}
where 
\begin{equation}
L[n]= \log{\left[\frac{\mu_s n}{mC_F\als}\right]}
\,,
\quad
S[a,n]=\sum_{k=1}^n \frac{1}{k^a}
\,,
\quad
S_2[a,b,n]=\sum_{k=1}^n \frac{1}{k^a}S[b,k]
\,.\end{equation}

For completeness we also give the corrections to the wave function
induced by the QED effect, \Eqn{EMpot}. They read
\begin{eqnarray}
\label{wfqed}
\delta \phi_n^{(s=1)} &\to& \delta \phi_n^{(s=1)} + 
\frac{3\, e_Q^2 \alpha_{\rm EM}}{C_F\, \alpha_s} 
+ \frac{e_Q^2 \alpha_{\rm EM}}{C_F\, \pi}
\left( \frac{3 a_1}{2} + 
\frac{3\pi \, e_Q^2 \alpha_{\rm EM}}{C_F\, \alpha^2_s} \right.
\\
&& \qquad  \qquad 
\left. +\ \beta_0 \left( 3 L[n]+S[1,n]+ 2 n S[2,n] - 
  \frac{5}{2} - \frac{n \pi^2}{3} \right)  \right) \, .
\nn
\end{eqnarray}
Of course, the corrections to the matching coefficients of the
current, \Eqns{c1qed}{c0qed} have to be taken into account as well.
Due to the additional suppression $e_Q^2 = 1/9$ these QED corrections
are numerically not very important.

From these expressions one can obtain the wave-function correction 
for the spin zero case
\begin{equation}
\delta \phi_n^{(s=0)}=\delta \phi_n^{(s=1)}+\delta \phi_n^{\Delta s}
\,,
\end{equation}
where
\begin{equation}
\delta \phi_n^{\Delta s}
=
-\frac{2}{3}C_F^2  D_{S^2,s}^{(2)} \als
\left(-2 L[n]+2 S[1,n]
-\frac{4}{n} -\frac{7}{3}\right)
\end{equation}
by using the results from Ref.~\cite{Penin:2004ay}. Therefore, the 
decay of the pseudoscalar heavy quarkonium to two photons reads
($d_0=d_1$)
\begin{eqnarray}
\label{pseudo}
\lefteqn{\Gamma(\eta_b(nS) \rightarrow \gamma\gamma) =} && \\
\nonumber && 
16 \pi\,  C_A \left[ { \alpha_{EM}\, e_Q^2 \over M_{\eta_b(nS)}} \right]^2
\left|\phi^{(s=0)}_n({\bf 0}) \right|^2
\left\{
c_0-d_0{M_{\eta_b(nS)}-2m \over 6m}
\right\}^2
\,.
\end{eqnarray}

We now perform a phenomenological analysis of these results.  We
restrict our analysis to the ground state of bottomonium. For the mass
of the $\eta_b(1S)$, we use the $\Upsilon(1S)$ mass, which is
consistent to the order of interest.  When we perform the numerical
analysis, we expand the expressions (except for the overall factor
$1/M^2_{\Upsilon(1S)/\eta_b(1S)}$) in \Eqns{vector}{pseudo}, and drop
subleading corrections, in order to work strictly at LL, NLL (NLO) and
NNLL (NNLO). In particular, this applies to the relativistic
correction proportional to $d_s$, where
$M_{\Upsilon(1S)/\eta_b(1S)}-2m$ is replaced by
$E_1^{(0)}=-mC_F^2\als^2/4$.  The expressions above have been written
in the pole scheme. Therefore, we change to a threshold scheme
suitable for the renormalon cancellation. We consider three possible
schemes: the PS, the RS and the RS' scheme with the subtraction scale
$\mu_F=2$~GeV. The numerical values of the bottom-quark mass to be
used in the various schemes are take from Ref.~\cite{Pineda:2006gx}
and are given by $m_{\PS}(2~{\rm GeV})=4.515$~GeV, $m_{\RS}(2\,{\rm
GeV})=4.370$~GeV, and $m_{\RS'}(2~{\rm GeV})=4.750$~GeV. Apart from
the different numerical values for $m$ to be used, the first change in
the formulae appears (at most) at NNLL (NNLO) and is due to the shift
\begin{eqnarray}
m &\rightarrow& m_X(\mu_f)+\delta m_X(\mu_f) \\
\left.\Gamma\right|_{m} &\rightarrow&
\left.\Gamma\right|_{m\rightarrow m_X}+
3\frac{\delta m_X}{m_X}\Gamma^{(0)}
\,,
\end{eqnarray}
where $m_X$ represents a generic threshold mass, $\Gamma^{(0)}$ is the
decay width at lowest order and the shift in the mass $\delta m_X \sim
mv^2$.  We notice that in the RS' scheme $\delta m_{\rm RS'}=0$ at
order $mv^2$. Therefore, in this scheme, with the current precision of
the calculation, the expressions for $\Gamma$ are equivalent to those
in the pole mass scheme.

\begin{figure}[h!]
   \begin{center}
    \includegraphics[width=.75\textwidth]{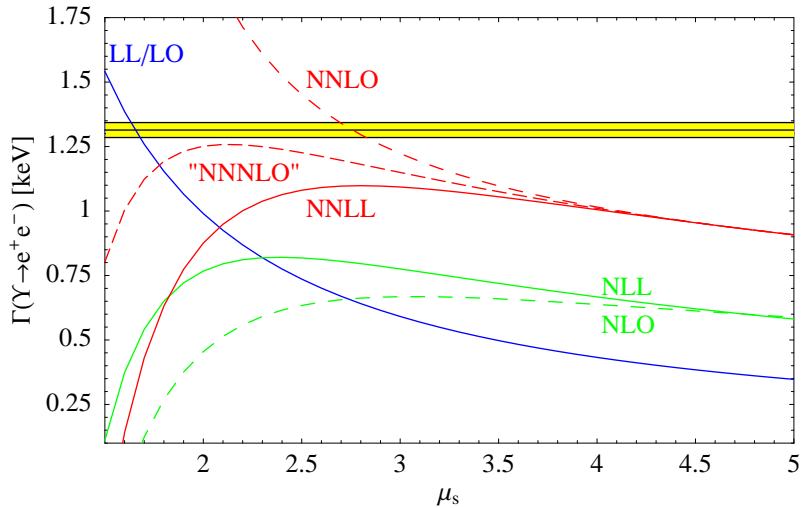}
   \caption{\small{Prediction for the $\Upsilon(1S)$ decay rate to
   $e^+e^-$.  We work in the RS' scheme.}}
   \label{fig:decayee}
   \end{center}
\end{figure}

\begin{figure}[h!]
   \begin{center}
    \includegraphics[width=.75\textwidth]{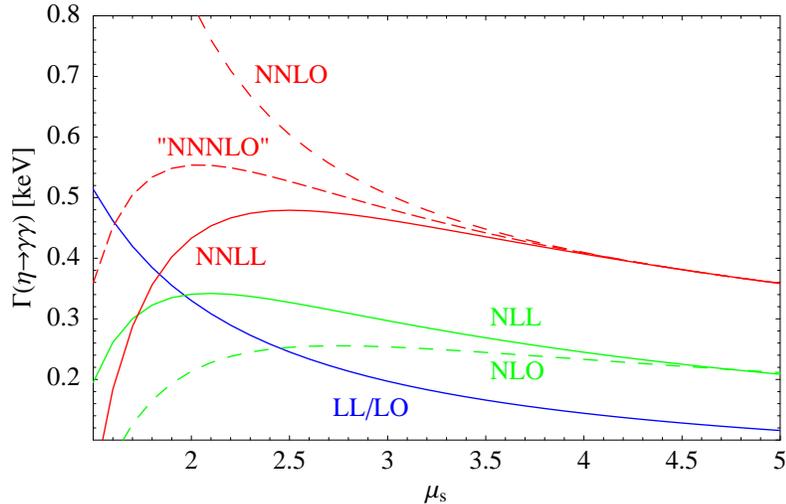}
   \caption{\small{Prediction for the $\eta_b(1S)$ decay rate to two
   photons.  We work in the RS' scheme.}}
   \label{fig:decaygg}
   \end{center}
\end{figure}

The results for the vector and pseudoscalar decay can be found in
Figure~\ref{fig:decayee} and Figure~\ref{fig:decaygg} respectively.
From the numerical analysis, we find that the NNLL corrections are
huge, especially for the $\eta_b(1S) \rightarrow \gamma\gamma$
decay. The result we obtain for this decay is compatible with the
number obtained in Ref. \cite{Penin:2004ay}. This is somewhat
reassuring, since in that reference the ratio of the spin-one
spin-zero decay was considered, which was much more scale independent,
as well as more convergent (yet still large) than for each of the
decays themselves.  This agreement can be traced back to the fact
that, for the spin-one decay, for which we can compare with
experiment, we find that the NNLL result improves the agreement with
the data.  Overall, the resummation of logarithms always significantly
improves over the NNLO result, the scale dependence greatly improves,
as well as the convergence of the series. On the other hand the
problem of lack of convergence of the perturbative series is not
really solved by the resummation of logarithms and it remains as an
open issue. Due to the lack of convergence we refrain from giving
numbers (and assigning errors) for our analysis. In this respect we
can not avoid to mention that, whereas the perturbative series in
non-relativistic sum rules is sign-alternating, is not
sign-alternating for the electromagnetic decays. Finally, we would
also like to remark the strong scale dependence that we observe at low
scales, which we believe to have the same origin than the one observed
in $t\bar{t}$ production near threshold in the previous section.

As for the $t\bar{t}$ case the ``NNNLO'' curve, which contains the
logarithmically enhanced NNNLO terms, follows closely the RGI curve up
to a certain value of $\mu_S$. For smaller soft scales, the two curves
start to deviate. However, in the $b\bar{b}$ case, this deviation
takes place already for reasonably large values of $\mu_s \sim
2.5$~GeV.  This seems to suggest that resummation is rather more
important in the $b\bar{b}$ case. However, we have to keep in mind
that some logarithms at NNLL are still missing in our analysis and
this might well affect this conclusion.

\begin{figure}[h!]
   \medskip
   \begin{center}
    \includegraphics[width=.75\textwidth]{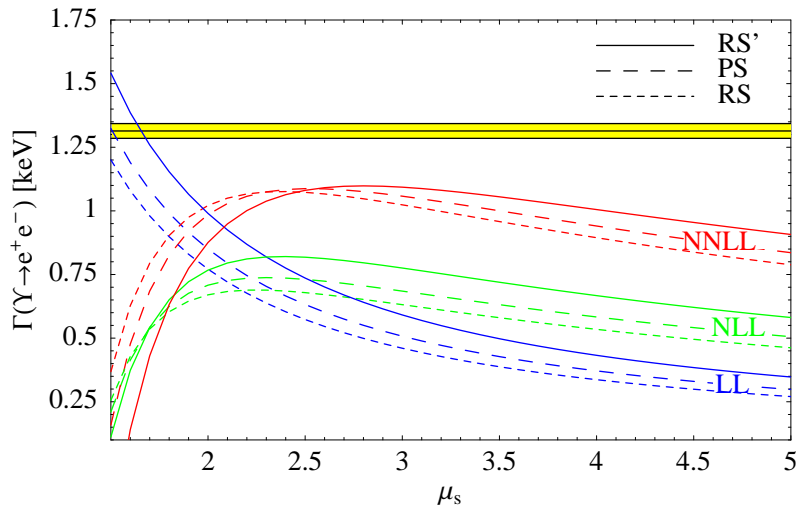}
   \caption{\small{Prediction for the $\Upsilon(1S)$ decay rate to
   $e^+e^-$ at LL, NLL and NNLL for the PS, RS and RS' mass.}}
   \label{fig:decayeethreshold}
   \end{center}
\end{figure}

We also illustrate the dependence on the threshold mass evaluation in
Figure~\ref{fig:decayeethreshold}. Compared with the uncertainty due to
the lack of convergence of the perturbative series, the dependence on
the threshold mass is negligible.

\section{Conclusions}

We have studied the effect of the resummation of logarithms for
$t\bar{t}$ production near threshold and inclusive electromagnetic
decays of heavy quarkonium.  This analysis is complete at NNLO and
includes the full resummation of logarithms at NLL accuracy and some
partial contributions at NNLL accuracy.

Compared with fixed-order computations the scale dependence and
convergence of the perturbative series is greatly improved for both
the position of the peak and the normalization of the total cross
section of $t\bar{t}$ production near threshold. Nevertheless, we
identify a possible source of a large scale dependence in the
result. While the result is very stable with respect to the variation
of the soft scale $\mu_s$ it shows a considerably larger dependence on
the hard scale $\mu_h$. This might well be related to the fact that
the variation with respect to $\mu_h$ is correlated with the ultrasoft
scale variation and the ultrasoft logarithms are not fully included at
the NNLL level in our result.  At present we estimate the remaining
theoretical uncertainty of the normalization of the total cross
section to be of the order of 10\% and for the position of the peak of
the order of 100 MeV, based on the difference between the NLL and NNLL
plots, as well as on the hard scale dependence. We note that this
estimate of the theoretical error is somewhat larger than in
Ref.~\cite{Hoang:2003ns}, due to the fact that we use more possible
sources of the theoretical error.  We would like to remark that the
use of the RGI provides us also with a significantly improved
determination of the top mass.

For the inclusive electromagnetic bottomonium decays the corrections
are even larger than in the top case. In the case of the $\Upsilon(1S)
\rightarrow e^+e^-$, they bring the final prediction into better
agreement with experiment. For the $\eta_b \rightarrow \gamma\gamma$,
there is no experimental data at present and the presence of huge
corrections make a reliable theoretical prediction difficult.

Finally we remark that the NNNLO part of the RGI results is
numerically considerably more important than the even higher-order
terms, in particular for the top case. Thus it would be highly
desirable to combine a full NNNLO evaluation with a complete NNLL
result in order to obtain a satisfactory theoretical prediction of
heavy-quark pair production near threshold.


\vspace{5mm}
\noindent
{\bf Acknowledgments:}\\ The work of A.P. was supported in part by a
{\it Distinci\'o} from the {\it Generalitat de Catalunya}, as well as
by the contracts MEC FPA2004-04582-C02-01, CIRIT 2005SGR-00564 and the
EU network EURIDICE, HPRN-CT2002-00311. A.P. acknowledges discussions
with A.A. Penin and J. Soto.


\appendix

\section{Insertions \label{AppIns}}

In this appendix we discuss how to obtain the perturbative expansion
of the Green function. For the loop integration measure we use 
\begin{equation}
\int d\tilde{p}_i \equiv  
\left(\frac{\mu^2 e^{\gamma_E}}{4\pi}\right)^{\epsilon} 
\int \frac{d^d {\bf p}_i}{(2\pi)^d}
\label{intmeas}
\end{equation}
such that the $\MS$-scheme corresponds to minimally subtracting the
poles in $\epsilon$.  We start by writing the potential,
\Eqn{Hpnrqcd}, for the case of a spin triplet S wave
\begin{eqnarray}
\delta \tilde V\Big|_{^3\!S_1} &=&
 -\,  \frac{C_F \alpha_s^2}{{\bf q}^2} 
    \left(a_1-\beta_0 \log\frac{{\bf q}^2}{\mu_s^2} \right)
	-\,  \frac{C_F \alpha_s^3}{{\bf q}^2}{C_A^3\over
  6\beta_0} \log\left[
\alpha_{s}(\mu_s)\over \alpha_{s}(\mu_{us}) \right]
\label{Hpnrqcd3S1}
\\
&&
- \, c_4 \frac{p^4}{4m^3}\, (2\pi)^d\delta^{(d)}({\bf q})
- \frac{2\pi C_F D^{(2)}_{1,s}}{m^2} \frac{{\bf p}^2+{\bf p}'^2}{{\bf q}^2}
\nonumber
\\
&&
- \,\frac{C_F \alpha_s^3}{4 \pi\, {\bf q}^2} 
  \left(a_2 -(2 a_1 \beta_0+\beta_1) \log\frac{{\bf q}^2}{\mu_s^2} +
        \beta_0^2 \log^2\frac{{\bf q}^2}{\mu_s^2} \right) 
\nonumber
\\
&& 
+ \, \frac{3\pi C_F D^{(2)}_{d,s}}{m^2}
- \frac{\pi C_F D^{(2)}_{S^2,s}}{m^2}\, 
       \frac{(d-4)(d-1)}{d}
\nonumber
\\
&& 
- \,\frac{\pi^2 C_F C_A D^{(1)}_s}{m\, q^{1+2\e}}\, (1-\e)
  \frac{e^{\e \gamma_E}}{\mu_s^{-2\e}}
  \frac{\Gamma^2(\frac{1}{2}-\e)\Gamma(\frac{1}{2}+\e)}
        {\pi^{3/2}\Gamma(1-2\e)}
\nonumber \\
&&
+ \, \frac{\pi C_F D^{(2)}_{2,s}}{m^2} \left(
      \left(\frac{{\bf p}^2-{\bf p}'^2}{{\bf q}^2}\right)^2 - 1 \right)
\nonumber 
\,,
\end{eqnarray}
where we have used the explicit expression for
$\alpha_{\tilde{V}_s}$. The terms in the first line of
\Eqn{Hpnrqcd3S1} are NLO and we have to consider double insertions of
such terms. All other terms are NNLO. Insertions of the potentials
with $D^{(2)}_{d,s}$, $D^{(2)}_{S^2,s}$, $D^{(1)}_{s}$ and
$D^{(2)}_{2,s}$ result in divergences. Thus, the corresponding
coefficients have to be known in $d$ dimensions. The insertions with
$c_4$, $D^{(2)}_{1,s}$ and $D^{(2)}_{2,s}$ can be related to other
insertions, discussed below, using the $d$-dimensional equation
\begin{eqnarray}
\lefteqn{
\left(\frac{{\bf p}^2}{m} - E\right) 
\tilde G_c({\bf p},{\bf p}^{\, \prime};E) = } &&
\label{eom} \\
&& (2\pi)^d \delta^{(d)} ({\bf p}-{\bf p}^{\, \prime}) 
+ \int \frac{d^d {\bf k}}{(2\pi)^d}
\frac{4 \pi C_F \alpha_s}{{\bf k}^2}
\tilde G_c({\bf p}-{\bf k},{\bf p}^{\, \prime};E) 
\nonumber
\,.
\end{eqnarray}
For the terms in the second line of \Eqn{Hpnrqcd3S1} we get
\begin{eqnarray}
\label{AppAline2}
\lefteqn{
\int \prod d\tilde{p}_i\ \tilde G_c({\bf p}_1,{\bf p}_2)\!
\left(\frac{{\bf p}_3^4 c_4}{4m^3}\, (2\pi)^d\delta^{(d)}({\bf q}_{23})
   + \frac{2\pi C_F D^{(2)}_{1,s}}{m^2} 
    \frac{{\bf p}_2^2+{\bf p}_3^2}{q_{23}^2} \right)
   \! \tilde G_c({\bf p}_3,{\bf p}_4)} &&
\\
&& = \ 
\frac{E}{2 m} c_4\,
\int \prod d\tilde{p}_i\ \tilde G_c({\bf p}_1,{\bf p}_2) \,
- \int \prod d\tilde{p}_i\ \tilde G_c({\bf p}_1,{\bf p}_2) \,
 \delta \tilde V_{\rm EOM}
 \, \tilde G_c({\bf p}_3,{\bf p}_4)
\nonumber
\end{eqnarray}
where we defined  ${\bf q}_{ij}\equiv {\bf p}_i - {\bf p}_j$, 
$q_{ij} \equiv |{\bf p}_i - {\bf p}_j|$ and
\begin{eqnarray}
\label{AppAVeom}
\delta \tilde V_{\rm EOM} &=& 
- c_4 \frac{E^2}{4m} (2\pi)^d\delta^{(d)}({\bf q}_{23})
- \left(2 c_4 \alpha_s + 4 D^{(2)}_{1,s}\right) 
  \pi C_F \frac{E}{m\, q_{23}^2}
\\
&& 
- \ \frac{C_F^2 \alpha_s \pi^2}{m\, q_{23}^{1+2\e}} 
\left( \alpha_s \frac{c_4}{2} + 2 D^{(2)}_{1,s} \right) 
 \frac{e^{\e \gamma_E}}{\mu_s^{-2\e}}
\frac{\Gamma^2(\frac{1}{2}-\e)\Gamma(\frac{1}{2}+\e)}
        {\pi^{3/2}\Gamma(1-2\e)}
\nonumber
\,.
\end{eqnarray}
We thus need the following insertions:
\begin{eqnarray}
{\cal I}_{l} &\equiv&
 \int \prod d\tilde{p}_i\ \tilde G_c({\bf p}_1,{\bf p}_2) \,
 \frac{\log[q_{23}^2/\mu_s^2]}{q_{23}^2} 
 \, \tilde G_c({\bf p}_3,{\bf p}_4) \\
{\cal I}_{l^2} &\equiv&
 \int \prod d\tilde{p}_i\ \tilde G_c({\bf p}_1,{\bf p}_2) \,
 \frac{\log^2[q_{23}^2/\mu_s^2]}{q_{23}^2}
 \, \tilde G_c({\bf p}_3,{\bf p}_4) \\
{\cal I}_{l,c} &\equiv&
 \int \prod d\tilde{p}_i\ \tilde G_c({\bf p}_1,{\bf p}_2) \,
 \frac{\log[q_{23}^2/\mu_s^2]}{q_{23}^2} 
 \, \tilde G_c({\bf p}_3,{\bf p}_4) \, 
 \frac{1}{q_{45}^2}\, \tilde G_c({\bf p}_5,{\bf p}_6) \\
{\cal I}_{l,l} &\equiv&
 \int \prod d\tilde{p}_i\ \tilde G_c({\bf p}_1,{\bf p}_2) \,
 \frac{\log[q_{23}^2/\mu_s^2]}{q_{23}^2} 
 \, \tilde G_c({\bf p}_3,{\bf p}_4) \, 
 \frac{\log[q_{45}^2/\mu_s^2]}{q_{45}^2} 
  \, \tilde G_c({\bf p}_5,{\bf p}_6) \\
{\cal I}_{1} &\equiv&
 \int \prod d\tilde{p}_i\ \tilde G_c({\bf p}_1,{\bf p}_2) 
 \, \tilde G_c({\bf p}_3,{\bf p}_4) \label{I3def} \\
{\cal I}_{\delta} &\equiv&
 \int \prod d\tilde{p}_i\ \tilde G_c({\bf p}_1,{\bf p}_2) \, 
 (2\pi)^{d-1}\delta({\bf p}_2-{\bf p}_3)
 \, \tilde G_c({\bf p}_3,{\bf p}_4) \label{I4def}  \\
{\cal I}_{na} &\equiv&
 \int \prod d\tilde{p}_i\ G({\bf p}_1,{\bf p}_2) \,
 \frac{\mu_s^{2\epsilon}}{q_{23}^{1+2\epsilon}}
 \, G({\bf p}_3,{\bf p}_4)  \label{I5def} 
\end{eqnarray}
All these insertions have been computed in
Ref.~\cite{Beneke:1999qg}. For the reader's convenience we present the
results here. We write them in terms of
\begin{eqnarray}
\ell &\equiv& \log\left[-\frac{4 m E}{\mu^2}\right] + 2 \bar{\psi}
     \equiv \log\left[-\frac{4 m E}{\mu^2}\right] + 2 \psi
              + 2 \gamma_E \label{AppAell}\\
{\cal H}_1  &\equiv& _3F_4(1,1,1,1;2,2,1-\lambda;1) \\
{\cal H}_2  &\equiv& _4F_5(1,1,1,1,1;2,2,2,1-\lambda;1)
\end{eqnarray}
and the argument of the $\psi$ functions ($\psi[x]\equiv d\log
\Gamma[x]/dx$) and its derivatives is always understood to be
$(1-\lambda) \equiv 1- (C_F \alpha_s)/(2 \sqrt{-E/m})$ unless
stated otherwise.

We start by writing the Green function before $\overline{\rm MS}$
subtraction
\begin{equation}
\label{AppAInsNo}
\tilde G_c(0,0;E) = \frac{C_F m^2 \alpha_s}{8 \pi} 
  \left( \frac{1}{2\epsilon} 
  - \ell - \frac{1}{\lambda}  + 1 
   + \epsilon\, {\cal G^\epsilon} \right)
\,,
\end{equation}
where ${\cal G^\epsilon}$ denotes the term of order $\epsilon$ of the
Green function. This term is not explicitly known, but it is not
needed for the calculation of the cross section, even though it results
in finite terms in some of the insertions below, because it cancels 
in the total sum.

The insertions of the higher-order corrections to the Coulomb
potential with terms of the form $\log^j[q^2/\mu_s^2]/q^2$ are
computed by taking derivatives with respect to $\kappa$ at $\kappa =
0$ of $\int \prod d\tilde{p}_i\ \tilde G_c({\bf p}_1,{\bf p}_2) \,
(q_{23}/\mu_s)^{-2-2\kappa} \, \tilde G_c({\bf p}_3,{\bf p}_4)$.
Taking the first derivative yields the insertion of a single
logarithm. We get
\begin{eqnarray}
\label{AppAI1Lb}
\left(\frac{m^2}{64 \pi^2}\right)^{-1} {\cal I}_{l} &=& 
\frac{1}{2\epsilon^2} + \frac{1}{\epsilon} - \ell^2 
+ 4\, \ell\, \lambda\, \psi' 
 \\
&+& 12\, \psi' - 4 \lambda\, \psi''
- 16\, {\cal H}_1 + 2 -\frac{\pi^2}{4}
\nonumber
\,.
\end{eqnarray}
The insertion of a logarithm squared is obtained by taking the second
derivative.
\begin{eqnarray}
\label{AppAI1L2b}
\left(\frac{m^2}{64 \pi^2}\right)^{-1} {\cal I}_{l^2} &=& 
\frac{1}{2\epsilon^3} + \frac{1}{\epsilon^2} 
+ \frac{2}{\epsilon} + \frac{\pi^2}{12\,\epsilon}
-\frac{2}{3} \ell^3 
+ 4\,\ell^2\, \lambda\, \psi'
\\
&-& 8\, \ell\, \Big( -3\, \psi'  +\ \lambda\, \psi'' 
+ 4\, {\cal H}_1
+ \frac{\pi^2}{12} \Big)
\nonumber \\
&-&  16 \lambda\, \psi'^2 
- 64 \lambda \Big(\bar{\psi}+\frac{1}{\lambda}-2-\frac{5\pi^2}{48}\Big)\,
\psi'
- 128 \,\bar{\psi}
\nonumber \\
&-& 32 \Big(\frac{2}{3} - \lambda\Big) \, \psi''
+ \frac{16}{3}\lambda\, \psi'''
- 128\, {\cal H}_2 - 64 \Big(1-\bar{\psi}\Big)\, {\cal H}_1
\nonumber \\
&+& 16\, S_1 + 52 \zeta(3)
+196 + \frac{11 \pi^2}{2}
\,,
\nonumber
\end{eqnarray}
where
\be
S_1=-\sum_{n=1}^{\infty}\frac{2\Gamma[n]\Gamma[1-\lambda]}
{(1+n)^2\Gamma[1+n-\lambda]}
\left(2\bar{\psi}[n]-2\bar{\psi}[1+n-\lambda]+(1+n)\psi'[1+n]\right)
\,.
\ee
Double insertions of Coulomb potentials are computed in a similar
way. The results read
\begin{equation}
\label{AppAI2Lb}
\left(\frac{m^2 \lambda}{64 \pi^3 C_F \alpha_s}\right)^{-1} 
{\cal I}_{l,c} =
\ell\, \Big(\psi' - \frac{\lambda}{2} \psi''\Big)
- 2 \bar{\psi}\, \psi'  + 2 \lambda\,\bar{\psi}\,\psi'' 
+ \frac{\lambda}{3}\,\psi''' + 2\, S_2
\,,
\end{equation}
where
\be
S_2=\sum_{n=1}^{\infty}
\frac{(n+\lambda)\bar{\psi}[n-\lambda]}{(n-\lambda)^3}
\,,
\ee
and 
\begin{eqnarray}
\label{AppAI2LLb}
\left(\frac{m^2 \lambda}{64 \pi^3 C_F \alpha_s}\right)^{-1} 
{\cal I}_{l,l} &=&
\ell^2\, \Big(\psi' - \frac{\lambda}{2} \psi'' \Big)
+ 4\, \ell\, \Big(\bar\psi\, (\lambda\,\psi'' - \psi') 
              + \frac{\lambda}{6}\, \psi'''+ S_2 \Big)
\\
&-& \frac{4}{\lambda}\, \bar\psi^3 - 8\lambda\,\bar\psi^2\,\psi''
- \frac{8}{\lambda^2} \,\bar\psi^2
\nonumber \\
&-& \frac{8}{\lambda}\,\bar\psi\,\psi' - 4\,\bar\psi\,\psi''
-\frac{8}{3}\lambda\,\bar\psi\,\psi'''  
- \frac{4}{\lambda^3}\, \bar\psi
- \frac{4}{\lambda^2}\,\psi'
\nonumber \\
&-&\frac{2}{\lambda}\,\psi'' - \frac{2}{3}\,\psi'''
-\frac{\lambda}{6}\,\psi''''
-4\, S_5 -8\, \bar{\psi}\, S_2
\,,
\nonumber
\end{eqnarray}
where
\be
S_5=\sum_{n=1}^{\infty}\frac{(n+\lambda)\bar\psi[n-\lambda]}
{(n-\lambda)^3}
\left(
-\frac{2\lambda}{n(n-\lambda)}
+\frac{2\lambda\bar\psi[1-\lambda]}{n}
-\frac{(n+\lambda)\bar\psi[n-\lambda]}{n}
\right)
\,.
\ee
Inserting 1 results in the square of the Green function
\begin{equation}
\label{AppAI3}
{\cal I}_1 = [G(0,0;E)]^2
\,.
\end{equation}
Note that ${\cal I}_1$ obtains a finite contribution due to ${\cal
  G^\epsilon}$.  The insertion due to the $\delta$ function can be
computed directly and  takes a very simple form
\begin{equation}
\label{AppAI4}
\left(\frac{m \lambda}{4 \pi C_F \alpha_s}\right)^{-1} 
{\cal I}_\delta = 2 \lambda^2 \psi' + 2 \lambda + 1
\,,
\end{equation}
while the insertion due to the non-analytic potential is more
complicated and, written in terms of
\begin{equation}
\label{AppALLdef}
\bar\ell \equiv \ell -3 + \log 2
\end{equation}
reads
\begin{eqnarray}
\left(\frac{m^3 C_F \alpha_s}{64 \pi^3}\right)^{-1} \! \!
{\cal I}_{na} &=& \frac{1}{2\epsilon^2} - \frac{1}{\epsilon}\, 
\Big( 2\, \bar\ell +\frac{2}{\lambda}+2-\log 2 \Big)
+ 2 \bar\ell^2 
\\
&+& \frac{4\, \bar\ell}{\lambda}
+ 2\, {\cal G^\epsilon} + 4 -\log^2 2
+ 4 \log 2 - \frac{5\pi^2}{24} 
\nonumber
\end{eqnarray}
Finally, the insertion due to the $L^2$ operator, \Eqn{lddef}, can be
related to ${\cal I}_{na}$ and ${\cal I}_1$ with the help of
\Eqn{l2relation} and we get
\begin{eqnarray}
{\cal I}_{L^2} &\equiv&
 \int \prod d\tilde{p}_i\ \tilde G_c({\bf p}_1,{\bf p}_2) \,
 \left(\left(\frac{{\bf p}_2^2-{\bf p}_3^2}{q_{23}^2}\right)^2 
  - 1 \right)
 \, \tilde G_c({\bf p}_3,{\bf p}_4) 
\label{AppAAMom}
\\
&=& \frac{C_F^2 m^4 \alpha_s^2}{2 (4\pi)^2} \left(
\frac{1}{4\e} - \ell - \frac{1}{2\lambda^2}-\frac{1}{\lambda}+3\right)
\nonumber
\end{eqnarray}
We are finally in a position to write down the expression for the
Green function
\begin{eqnarray}
\label{AppAGreen}
\lefteqn{\tilde G_{s=1}^{\rm NNLL} =} &&
\\ &&
\left(1-\frac{c_4\, E}{2m}\right)
\tilde G_c\Big|_{\alpha_s \to \alpha_s \left(1+\frac{E}{2m} c_4
  +\frac{E}{m} \frac{D^{(2)}_{1,s}}{\alpha_s}
  + \frac{\alpha_s}{4\pi} a_1
  + \frac{e_Q^2 \alpha_{\rm EM}}{C_F \alpha_s}
  + \frac{\alpha^2_s}{(4\pi)^2} a_2 +{C_A^3\over
  6\beta_0}\als^2 \log\left[
\alpha_{s}(\mu_s)\over \alpha_{s}(\mu_{us}) \right]\right) } 
\nonumber
\\
&& + \ i^2 C_F \left( \alpha_s^2 \beta_0 
  + \frac{\alpha^3_s}{4\pi} (2 a_1 \beta_0 + \beta_1) \right) 
    {\cal I}_l
 - i^2 \frac{\alpha^3_s}{4\pi} C_F \beta_0^2 \  {\cal I}_{l^2}
\nonumber \\
&&
-\ i^4 2 \alpha_s^4 \beta_0 C_F^2 \left(a_1 + 
      \frac{4\pi e_Q^2}{C_F} \frac{\alpha_{\rm EM}}{\alpha^2_s}\right)\,  
    {\cal I}_{l,c}
+ i^4 \alpha_s^4 C_F^2 \beta^2_0 \, {\cal I}_{l,l}
- i^2 \frac{E^2}{4 m} c_4\, {\cal I}_\delta
\nonumber \\
&& +\  i^2 \frac{\pi C_F}{m^2} D^{(2)}_{2,s}\, {\cal I}_{L^2}
+ i^2 \frac{\pi C_F}{m^2}\left(
  3 D^{(2)}_{d,s} +  D^{(2)}_{S^2,s} \frac{(d-4)(d-1)}{d} \right)\, 
  {\cal I}_1
\nonumber \\
&& -\ i^2 \left( \frac{C_F^2 \alpha_s \pi^2}{2 m} 
   \left(c_4 \alpha_s + 4 D^{(2)}_{1,s}\right) 
  + \frac{C_A C_F \pi^2}{m} (1-\e) D^{(1)}_s \right) 
\nonumber \\
&& \hspace*{3cm} \times \  e^{\e \gamma_E}\, 
   \frac{\Gamma^2(\frac{1}{2}-\e)\Gamma(\frac{1}{2}+\e)}
        {  \pi^{3/2}\Gamma(1-2\e)}\,   {\cal I}_{na}
\nonumber 
\,.
\end{eqnarray}
We mention once more that we use strictly expanded results. Thus the
terms beyond NNLL that are present in \Eqn{AppAGreen} are dropped.

\section{RGI potentials}

For ease of reference we explicitly display the RGI potentials not
shown in the main body of the paper (we remind that
$\mu_{us}=\mu_s^2/\mu_h$):

\beqa
{\alpha}_{{\tilde V}_s}(\mu_s) &=&\alpha_{\rm s}(\mu_s)
\left\{1+
\left(a_1-\beta_0 \log\frac{{\bf q}^2}{\mu_s^2} \right)
 {\alpha_{\rm s}(\mu_s) \over 4\pi}\right.
\nonumber\\
&&
\left.
\left(a_2 -(2 a_1 \beta_0+\beta_1) \log\frac{{\bf q}^2}{\mu_s^2} +
        \beta_0^2 \log^2\frac{{\bf q}^2}{\mu_s^2} \right)  
		{\alpha_{\rm s}^2(\mu_s) \over 16\,\pi^2}
 \right\}
\nn
\\
&&
+
{C_A^3\over
  6\beta_0}\als^3(\mu_s) \log\left[
\alpha_{s}(\mu_s)\over \alpha_{s}(\mu_{us}) \right],\\
D^{(1)}_s(\mu_{s})&=&\als^2(\mu_s)\left\{1+
{16\over 3\beta_0}\left({C_A \over 2}+C_F \right)
\log\left[
\alpha_{s}(\mu_s)\over \alpha_{s}(\mu_{us}) \right]\right\},
\label{Ds1}
\nn\\ 
D^{(2)}_{1,s}(\mu_{s})&=&\alpha_{\rm s}(\mu_s)\left\{1+
{8C_A\over 3\beta_0}\log\left[
\alpha_{s}(\mu_s)\over \alpha_{s}(\mu_{us}) \right]\right\},
\label{Ds2}
\nn
\\ 
D^{(2)}_{2,s}(\mu_{s})&=&\alpha_{\rm s}(\mu_s),
\label{Ds22}
\nn
\\ 
D^{(2)}_{S^2,s}(\mu_s)&=&\alpha_{\rm s}(\mu_s)c_F^2(\mu_s) - {3 \over 2\pi
C_F}(d_{sv}(\mu_s)+C_F d_{vv}(\mu_s)) ,
\label{Dss2i}\nn\\ 
D^{(2)}_{LS,s}(\mu_s)&=& {\alpha_{\rm s}(\mu_s) \over 3}
(c_S(\mu_s)+2c_F(\mu_s)), 
\label{DLs2i}\nn\\ 
D^{(2)}_{S_{12},s}(\mu_s)&=&\alpha_{\rm s}(\mu_s) c_F^2(\mu_s),
\label{Dsten2i}\nn
\eeqa
where ($z=\left[{\als(\mu_s) \over \als(\mu_h)}\right]^{1 \over
\beta_0}\simeq 1 -1/(2\pi)\als(\mu_s)\log ({\mu_s \over \mu_h})$,
$\beta_0={11 \over 3}C_A -{4 \over 3}T_Fn_f$)
\beqa
c_F(\mu_s)&=&z^{-C_A}
\,,
\nn\\
c_S(\mu_s)&=&2z^{-C_A}-1
\,,
\nn\\
c_D(\mu_s)&=&
{9C_A \over 9C_A+8T_Fn_f}
\left\{
-\frac{5 C_A + 4 T_F n_f}{4 C_A + 4
T_F n_f} z^{-2 C_A} +
\frac{C_A +16 C_F - 8 T_F n_f}{2(C_A-2T_F n_f)}
\right.
\nn
\\
&&\qquad
+ \frac{-7 C_A^2 + 32 C_A C_F - 4 C_A T_F n_f +32 C_F T_F
n_f}{4(C_A + T_F n_f)(2 T_F n_f-C_A)} z^{4 T_F n_f/3 - 2C_A/3}
\nn
\\
&&
\qquad
\left.
+{8T_Fn_f \over 9C_A}
\left[
z^{-2C_A}+\left({20 \over 13}+{32 \over 13}{C_F \over
C_A}\right)\left[1-z^{-13C_A \over 6}\right]
\right]
\right\}
\,,
\nn\\
\frac{d_{ss}(\mu_s)}{C_F}+d_{vs}(\mu_s)&=&
-\left(2C_F-3C_A\right){2\pi\over\beta_0}\als(\mu_h)
\left[z^{\beta_0}-1 \right]
\nn
\\
&&
-{27C_A^2 \over 9C_A+8T_Fn_f}{\pi \over \beta_0}\als(\mu_h)
\left\{
-\frac{5 C_A + 4 T_F n_f}{4 C_A + 4
T_F n_f}{\beta_0 \over \beta_0-2C_A}\left(z^{\beta_0-2 C_A}-1\right) 
\right.
\nn
\\
&&
\qquad
+
\frac{C_A +16 C_F - 8 T_F n_f}{2(C_A-2T_F n_f)}\left(z^{\beta_0}-1\right)
\nn
\\
&&
\qquad
+ \frac{-7 C_A^2 + 32 C_A C_F - 4 C_A T_F n_f +32 C_F T_F
n_f}{4(C_A + T_F n_f)(2 T_F n_f-C_A)}
\nn
\\
&&
\qquad
\qquad
\times
{3\beta_0 \over 3\beta_0+4T_Fn_f-2C_A}
\left( z^{\beta_0+4 T_F n_f/3 - 2C_A/3}-1\right)
\nn
\\
&&
\qquad
+{8T_Fn_f \over 9C_A}
\left[{\beta_0 \over \beta_0-2C_A}\left(z^{\beta_0-2C_A}-1\right)
+\left({20 \over 13}+{32 \over 13}{C_F \over C_A}\right)
\right.
\nn
\\
&&
\qquad\qquad
\left.
\left.
\times
\left(
\left[z^{\beta_0}-1\right]-{6\beta_0 \over 6\beta_0-13C_A}
\left[z^{\beta_0-{13C_A \over 6}}-1\right]
\right)
\right]
\right\}
\,,
\nn\\
\frac{d_{sv}(\mu_s)}{C_F}+ d_{vv}(\mu_s)
&=&{C_A \over
\beta_0-2C_A}\pi\als(\mu_h)\left\{z^{\beta_0-2C_A}-1\right\}
\label{RGeqhs}
\,.
\eeqa



\begin{thebibliography}{99}

\bibitem{FadinKhoze}
V.~S.~Fadin and V.~A.~Khoze,
JETP Lett.\  {\bf 46}, 525 (1987)
[Pisma Zh.\ Eksp.\ Teor.\ Fiz.\  {\bf 46}, 417 (1987)]; \\
V.~S.~Fadin and V.~A.~Khoze,
Sov.\ J.\ Nucl.\ Phys.\  {\bf 48}, 309 (1988)
[Yad.\ Fiz.\  {\bf 48}, 487 (1988)].

\bibitem{Martinez:2002st}
M.~Martinez and R.~Miquel,
Eur.\ Phys.\ J.\ C {\bf 27}, 49 (2003)
[arXiv:hep-ph/0207315].

\bibitem{Hoang:2000yr}
A.~H.~Hoang {\it et al.},
Eur.\ Phys.\ J.\ directC {\bf 2}, 1 (2000)
[arXiv:hep-ph/0001286].

\bibitem{bottomSR}
V.~A.~Novikov {\it et al.},
Phys.\ Rev.\ Lett.\  {\bf 38}, 626 (1977)
[Erratum-ibid.\  {\bf 38}, 791 (1977)]; 
V.~A.~Novikov {\it et al.},
Phys.\ Rept.\  {\bf 41}, 1 (1978).


\bibitem{Melnikov2} K. Melnikov and A. Yelkhovsky, Phys. Rev. {\bf
  D59} 114009 (1999). 

\bibitem{Penin} 
A.~A.~Penin and A.~A.~Pivovarov,
Nucl.\ Phys.\ B {\bf 549}, 217 (1999)
[arXiv:hep-ph/9807421].

\bibitem{Hoang:1999ye}
A.~H.~Hoang,
Phys.\ Rev.\ D {\bf 59}, 014039 (1999)
[arXiv:hep-ph/9803454]; \\
A.~H.~Hoang,
Phys.\ Rev.\ D {\bf 61}, 034005 (2000)
[arXiv:hep-ph/9905550].

\bibitem{Beneke:1999fe}
M.~Beneke and A.~Signer,
Phys.\ Lett.\ B {\bf 471} (1999) 233
[arXiv:hep-ph/9906475].


\bibitem{Pineda:2006gx}
A.~Pineda and A.~Signer,
Phys.\ Rev.\ D {\bf 73}, 111501 (2006)
[arXiv:hep-ph/0601185].

\bibitem{Brambilla:2004jw}
N.~Brambilla, A.~Pineda, J.~Soto and A.~Vairo,
Rev.\ Mod.\ Phys.\  {\bf 77}, 1423 (2005).

\bibitem{Beneke:1997zp}
M.~Beneke and V.~A.~Smirnov,
Nucl.\ Phys.\ B {\bf 522}, 321 (1998)
[arXiv:hep-ph/9711391].

\bibitem{nrqcd}
W.~E.~Caswell and G.~P.~Lepage,
Phys.\ Lett.\ B {\bf 167}, 437 (1986).


\bibitem{pnrqcd}
A.~Pineda and J.~Soto,
Nucl.\ Phys.\ Proc.\ Suppl.\  {\bf 64}, 428 (1998)
[arXiv:hep-ph/9707481]; \\
A.~Pineda and J.~Soto,
Phys.\ Rev.\ D {\bf 59}, 016005 (1999)
[arXiv:hep-ph/9805424].

\bibitem{Brambilla:1999xf}
  N.~Brambilla, A.~Pineda, J.~Soto and A.~Vairo,
  Nucl.\ Phys.\ B {\bf 566}, 275 (2000)
  [arXiv:hep-ph/9907240].

\bibitem{Brambilla:1999qa}
N.~Brambilla, A.~Pineda, J.~Soto and A.~Vairo,
Phys.\ Rev.\ D {\bf 60}, 091502 (1999)
[arXiv:hep-ph/9903355].

\bibitem{Kniehl:1999ud}
B.~A.~Kniehl and A.~A.~Penin,
Nucl.\ Phys.\ B {\bf 563}, 200 (1999)
[arXiv:hep-ph/9907489]; \\
B.~A.~Kniehl and A.~A.~Penin,
Nucl.\ Phys.\ B {\bf 577}, 197 (2000)
[arXiv:hep-ph/9911414].

\bibitem{Brambilla:1999xj}
 N.~Brambilla, A.~Pineda, J.~Soto and A.~Vairo,
Phys.\ Lett.\ B {\bf 470}, 215 (1999) 
[arXiv:hep-ph/9910238].

\bibitem{Kniehl:2001ju}
B.~A.~Kniehl, A.~A.~Penin, M.~Steinhauser and V.~A.~Smirnov,
Phys.\ Rev.\ D {\bf 65}, 091503 (2002)
[arXiv:hep-ph/0106135].

\bibitem{Penin:2005eu}
A.~A.~Penin, V.~A.~Smirnov and M.~Steinhauser,
Nucl.\ Phys.\ B {\bf 716}, 303 (2005)
[arXiv:hep-ph/0501042].

\bibitem{Beneke:2005hg}
M.~Beneke, Y.~Kiyo and K.~Schuller,
Nucl.\ Phys.\ B {\bf 714}, 67 (2005)
[arXiv:hep-ph/0501289].

\bibitem{Marquard:2006qi}
P.~Marquard, J.~H.~Piclum, D.~Seidel and M.~Steinhauser,
arXiv:hep-ph/0607168.

\bibitem{Pineda:2001ra}
A.~Pineda,
Phys.\ Rev.\ D {\bf 65}, 074007 (2002)
[arXiv:hep-ph/0109117].

\bibitem{Hoang:2002yy}
A.~H.~Hoang and I.~W.~Stewart,
Phys.\ Rev.\ D {\bf 67}, 114020 (2003)
[arXiv:hep-ph/0209340].

\bibitem{Kniehl:2003ap}
B.~A.~Kniehl, A.~A.~Penin, A.~Pineda, V.~A.~Smirnov and M.~Steinhauser,
Phys.\ Rev.\ Lett.\  {\bf 92}, 242001 (2004)
[arXiv:hep-ph/0312086].


\bibitem{Penin:2004xi}
A.~A.~Penin, A.~Pineda, V.~A.~Smirnov and M.~Steinhauser,
Phys.\ Lett.\ B {\bf 593}, 124 (2004)
[arXiv:hep-ph/0403080].

\bibitem{Pineda:2001et}
A.~Pineda,
Phys.\ Rev.\ D {\bf 66} (2002) 054022
[arXiv:hep-ph/0110216].

\bibitem{Penin:2004ay}
A.~A.~Penin, A.~Pineda, V.~A.~Smirnov and M.~Steinhauser,
Nucl.\ Phys.\ B {\bf 699}, 183 (2004)
[arXiv:hep-ph/0406175].

\bibitem{Hoang:2000ib}
A.~H.~Hoang, A.~V.~Manohar, I.~W.~Stewart and T.~Teubner,
Phys.\ Rev.\ Lett.\  {\bf 86}, 1951 (2001)
[arXiv:hep-ph/0011254].

\bibitem{Hoang:2001mm}
A.~H.~Hoang, A.~V.~Manohar, I.~W.~Stewart and T.~Teubner,
Phys.\ Rev.\ D {\bf 65}, 014014 (2002)
[arXiv:hep-ph/0107144].


\bibitem{vnrqcd}
M.~E.~Luke, A.~V.~Manohar and I.~Z.~Rothstein,
Phys.\ Rev.\ D {\bf 61}, 074025 (2000)
[arXiv:hep-ph/9910209]. 

\bibitem{massdef}
I.~I.~Y.~Bigi, M.~A.~Shifman and N.~Uraltsev,
Ann.\ Rev.\ Nucl.\ Part.\ Sci.\  {\bf 47}, 591 (1997)
[arXiv:hep-ph/9703290]; \\
A.~H.~Hoang, Z.~Ligeti and A.~V.~Manohar,
Phys.\ Rev.\ Lett.\  {\bf 82}, 277 (1999)
[arXiv:hep-ph/9809423].

\bibitem{Beneke:1998rk}
M.~Beneke,
Phys.\ Lett.\ B {\bf 434}, 115 (1998)
[arXiv:hep-ph/9804241].

\bibitem{Pineda:2001zq}
A.~Pineda,
JHEP {\bf 0106} (2001) 022
[arXiv:hep-ph/0105008].

\bibitem{Beneke:1999qg}
M.~Beneke, A.~Signer and V.~A.~Smirnov,
Phys.\ Lett.\ B {\bf 454}, 137 (1999)
[arXiv:hep-ph/9903260].

\bibitem{Beneke:1999zr}
M.~Beneke,
arXiv:hep-ph/9911490.

\bibitem{c1nnlo}
A.~Czarnecki and K.~Melnikov,
Phys.\ Rev.\ Lett.\  {\bf 80}, 2531 (1998)
[arXiv:hep-ph/9712222]; \\
M.~Beneke, A.~Signer and V.~A.~Smirnov,
Phys.\ Rev.\ Lett.\  {\bf 80}, 2535 (1998)
[arXiv:hep-ph/9712302].

\bibitem{Czarnecki:2001zc}
A.~Czarnecki and K.~Melnikov,
Phys.\ Lett.\ B {\bf 519}, 212 (2001)
[arXiv:hep-ph/0109054].

\bibitem{static}
Y.~Schroder,
Phys.\ Lett.\ B {\bf 447}, 321 (1999)
[arXiv:hep-ph/9812205];\\
M.~Peter,
Phys.\ Rev.\ Lett.\  {\bf 78}, 602 (1997)
[arXiv:hep-ph/9610209]. 

\bibitem{Pineda:2000gz}
A.~Pineda and J.~Soto,
Phys.\ Lett.\ B {\bf 495}, 323 (2000)
[arXiv:hep-ph/0007197].

\bibitem{Bodwin:1994jh}
G.~T.~Bodwin, E.~Braaten and G.~P.~Lepage,
Phys.\ Rev.\ D {\bf 51}, 1125 (1995)
[Erratum-ibid.\ D {\bf 55}, 5853 (1997)]
[arXiv:hep-ph/9407339].

\bibitem{Brambilla:2002nu}
N.~Brambilla, D.~Eiras, A.~Pineda, J.~Soto and A.~Vairo,
Phys.\ Rev.\ D {\bf 67}, 034018 (2003)
[arXiv:hep-ph/0208019].

\bibitem{upet}
A.~P.~Chapovsky, V.~A.~Khoze, A.~Signer and W.~J.~Stirling,
Nucl.\ Phys.\ B {\bf 621}, 257 (2002)
[arXiv:hep-ph/0108190]; \\
M.~Beneke, A.~P.~Chapovsky, A.~Signer and G.~Zanderighi,
Phys.\ Rev.\ Lett.\  {\bf 93}, 011602 (2004)
[arXiv:hep-ph/0312331]; \\
M.~Beneke, A.~P.~Chapovsky, A.~Signer and G.~Zanderighi,
Nucl.\ Phys.\ B {\bf 686}, 205 (2004)
[arXiv:hep-ph/0401002].

\bibitem{nfc}
V.~S.~Fadin, V.~A.~Khoze and A.~D.~Martin,
Phys.\ Rev.\ D {\bf 49}, 2247 (1994); \\
V.~S.~Fadin, V.~A.~Khoze and A.~D.~Martin,
Phys.\ Lett.\ B {\bf 320}, 141 (1994)
[arXiv:hep-ph/9309234]; \\
K.~Melnikov and O.~I.~Yakovlev,
Phys.\ Lett.\ B {\bf 324}, 217 (1994)
[arXiv:hep-ph/9302311].


\bibitem{Hoang:2004tg}
A.~H.~Hoang and C.~J.~Reisser,
Phys.\ Rev.\ D {\bf 71}, 074022 (2005)
[arXiv:hep-ph/0412258].

\bibitem{Hoang:2003ns}
A.~H.~Hoang,
Phys.\ Rev.\ D {\bf 69}, 034009 (2004)
[arXiv:hep-ph/0307376].


  
\end{thebibliography}
\end{document}